\documentclass[a4paper,onecolumn,11pt,accepted=2026-05-26]{quantumarticle}
\pdfoutput=1

\usepackage[utf8]{inputenc}
\usepackage[english]{babel}
\usepackage[T1]{fontenc}

\usepackage{amsmath,amssymb,amsfonts,mathrsfs,mathtools}
\usepackage{multirow,rotating,float}

\usepackage{hyperref}
\usepackage[numbers,sort&compress]{natbib}

\begin{document}

\title{Bosonic content of three-fermion highest-spin states}

\author{Jerzy Cioslowski}
\affiliation{Institute of Physics, University of Szczecin, Wielkopolska 15, 70-451 Szczecin, Poland}
\email{jerzy@wmf.univ.szczecin.pl}
\orcid{0000-0002-3713-9554}
\author{Krzysztof Strasburger}
\affiliation{Department of Physical and Quantum Chemistry, Faculty of Chemistry, Wrocław University of Science and Technology, Wybrzeże Wyspiańskiego 27, 50-370 Wrocław, Poland}
\email{krzysztof.strasburger@pwr.edu.pl}
\orcid{0000-0002-9214-7282}
\author{Denis K. Sunko}
\affiliation{Department of Physics, University of Zagreb Faculty of Science, Bijenička cesta 32, 10000 Zagreb, Croatia}
\email{dks@phy.hr}
\orcid{0000-0002-1383-0674}
\maketitle

\begin{abstract}
A rigorous characterization of the information content of any highest-spin three-fermion wave function is presented. It is based upon a formal decomposition of the wave function into a finite set of fixed invariants, called shapes, whose sole purpose is to satisfy the Pauli principle, and a variable part, constituting the bosonic excitations of these invariants, that provides its physical content. As an example, this decomposition is applied to a benchmark-quality approximate wave function of the lowest-energy quartet electronic state of the lithium atom. This wave function, which comprises hundreds of basis functions, is reduced to eleven shape blocks, only five of which are numerically significant. Such a compact characterization is a generic example of the appearance of superselection rules in configuration space, and provides a qualitative aid in the search for robust few-particle entangled states.
\end{abstract}

\section{Introduction}\label{introduction}

Powered by mutually reinforcing advances in algorithms involving entangled quantum states and technological abilities to implement them~\cite{Jouzdani24,Herrmann22}, research on quantum information, computing, and signal-processing has experienced remarkable progress over the past two decades~\cite{Pavarini21}. However, whereas abstract algorithms operate at the level of quantum numbers, device-level implementations are concerned with the actual photons and electrons whose wave functions have the required abstract properties. The maintenance of quantum coherence remains a major issue when those two meet at the implementation level~\cite{Saxena20}.

One of the differences between photons and electrons is that photons are charge-neutral, while electrons couple to each other by the Coulomb force. Hence, photon–photon interactions are negligible and any dissipative interaction is mediated by the environment, while the internal dynamics of electronic states needs to be considered from the outset in quantum design. The usual spin-qubit implementations manipulate information in the spin sector, while the spatial structure of the wave function typically controls relaxation via energy gaps, selection rules, and/or coupling to the environment, thus affecting coherence times. Therefore, robustness of the latter is a prime design consideration. A robust state is one that does not deexcite easily, typically because it would require a significant reconfiguration of the wave function in space.

In this work, we develop a formalism for the analysis of the spatial component of an arbitrary highest-spin three-fermion wave function in terms of fundamental many-body invariants of the Pauli principle called \emph{shapes}~\cite{Sunko16-1}, and apply it to a non-trivial realistic example. The method is based on Hilbert's ground-breaking result that invariants of all the classical and finite groups are finitely generated~\cite{Sturmfels08}. In our context, it means that there is only a finite number of intrinsically different ways a many-body wave function can satisfy the Pauli principle. A pedagogical introduction to shapes from the perspective of invariant theory is found in Ref.~\cite{Sunko22}. In particular, there are $36$ distinct shapes for $3$ fermions. Furthermore, these $36$ shapes fall into only $11$ classes according to the particular ways their antisymmetries split along the lower spatial dimensions. For example, one such class comprises four shapes that are all antisymmetric in the $xy$ plane and constant in the $z$ direction, another one encompasses those symmetric in a single plane and antisymmetric in the orthogonal directions, etc.

The importance of this scheme is twofold. First, satisfying the Pauli principle is by itself an elementary kinematic requirement, like satisfying the conservation of momentum. Our approach cleanly separates obligatory fulfillment of this kinematic requirement from the freedom inherent in the dynamics. The latter is manifested as symmetric functions that are a mathematically precise implementation of the physical notion of bosonic excitations of the shapes, i.e., a formal algebraic definition of density oscillations. We develop explicit formulas for the extraction of these excitations from a general three-fermion highest-spin wave function.

Second, and building on the first, kinematics dominates dynamics physically. Hence, kinematically distinct wave functions are expected to be robust when interacting, much more so than if one could turn one into the other by simply shedding a bosonic excitation (for example, orthohydrogen and parahydrogen cannot turn into each other by scattering, so they behave like a mixture of two different gases at arbitrary concentration; they are superselected in the sense that an entangled superposition is not possible because the amplitude for tunnelling between them is zero). Thus, the spatial component of the wave function is a universal source of practical superselection rules in configuration space. The aforementioned $11$ classes are the three-electron generalization of this ortho/parahydrogen paradigm, and a generic working example of using shapes as an organizational principle in that space.

In three dimensions, shapes are polynomials in $3N$ variables that can all be obtained as derivatives of a single predecessor called the source shape. They are also interesting from the information-theoretic point of view~\cite{Aliverti-Piuri24}, because this derivative structure allows an intrinsic information content (absolute entropy) to be defined for them, although they are pure states whose von Neumann entropy is zero~\cite{Sunko22}. This definition is based upon the simple observation that taking the derivative of a polynomial reduces its information content. The resulting information measure is an aggregate statistic related to the nodal structure of the many-body wave function.

The organization of the present paper is as follows: First, a brief overview of shapes is presented. Next, the pertinent formalism and main expressions for the analysis of an arbitrary three-fermion wave function in terms of shapes and concomitant bosonic excitations are derived. A group-theoretic classification of shapes is found in Appendix~\ref{altclass}. In the second part, a qualitative analysis of a benchmark-quality approximate wave function of the lowest-energy highest-spin electronic state of the lithium atom is presented. This wave function is decomposed into products of shapes and respective bosonic excitations. Thus, such excitations of a fermion system are identified without any theoretical arbitrariness for the first time. The paper concludes with a discussion placing these results in a wider context, including open issues.

\section{Shapes}\label{shapes}

The realization that many-fermion wave functions must be antisymmetric is formally encoded in Heisenberg's standard basis~\cite{Heisenberg26}, commonly known as Slater determinants~\cite{Slater29}. Mathematically, these are \emph{primitive realizations} of the Pauli principle: using them in calculations is a formal safeguard against violating it. A detailed description of the various operational representations of many-body fermion functions is found in Ref.~\cite{Koch21}.

It appears now that the many-body Hilbert space has a finer structure in addition to being a vector space~\cite{Sunko16-1}. The spatial component of a wave function of $N$ identical particles (a wave function for short) in $d$ dimensions,
\[\Psi := \Psi(\mathbf{r}_1, \ldots, \mathbf{r}_N) := 
\Psi(x_1, \ldots, x_N, y_1, \ldots, y_N,\ldots),\]
can be presented as a free module,
\begin{equation}
\Psi=\sum_{i=1}^D\Phi_iS_i,
\label{scheme}
\end{equation} 
generated by a finite number $D=(N!)^{d-1}$ of linearly independent functions $S_i:=S_i(\mathbf{r}_1,\allowbreak \ldots, \mathbf{r}_N)$ in $d$ sets (space dimensions) of $N$ variables $x_1, \ldots, x_N$, etc.,where the functions $\Phi_i:=\Phi_i(\mathbf{r}_1, \ldots, \mathbf{r}_N)$ are linear combinations of terms symmetric in each of these sets separately (thus called bosonic functions throughout this paper). This choice of coefficient ring is physically motivated by the nature of density waves in many-body systems that can in principle be resolved into independent plane waves in the three space directions. Each plane wave is a boson (a symmetric function) by itself, so the structure of a general physical bosonic excitation is less general than of an arbitrary symmetric function. The simplest way to interpret the free module is as a relaxation of the condition that the coefficients in the usual (infinite) expansion in Slater determinants are $c$-numbers. If they are allowed to be symmetric functions instead, only a finite number of antisymmetric basis functions is needed. These generators of the free module are called \emph{shapes}.

It should be noted that the term ``bosonic'' employed in this paper refers to the above mathematical symmetry property of the coefficients, rather than to any real (particle) system. The Hilbert space of real systems composed of physical bosons can also be presented in the form~\eqref{scheme}, with the same bosonic-function coefficent ring, but the corresponding shapes are symmetric functions that are different than either the bosonic functions or the antisymmetric fermion shapes~\cite{Sunko16-1}.

Unlike the more familiar vector-space approach to many-body Hilbert space, the free-module approach is inherently nonlinear. Shapes are geometric objects in the wave-function space that restrict fermion motion kinematically. They provide qualitative insights into the many-body wave function, e.g.\ bands in the spectra of finite systems are interpreted as ideals generated by their band-heads~\cite{Rozman20}. In this way, algebraic geometry becomes the natural framework for the particle picture of quantum mechanics~\cite{Sunko20}. Because of this additional structural insight, one can rewrite an arbitrary antisymmetric function in $\mathrm{L}^2(\mathbb{R}^3)^N$ in a much more compact form, as it is shown in the numerical example below.

The square-integrability of $\Psi$ is retained in the representation~\eqref{scheme}, the exponentially decaying terms being transferred to the functions $\Phi_i$. Slater determinants of analytic functions in a single variable are always divisible by an antisymmetric polynomial, so multiplying and dividing by the same polynomial brings us to the form~\eqref{scheme}. (If the argument refers to more than one particle, it has to be symmetric with respect to their exchange, because localization factors do not have good parity.) The physical question is whether the expression~\eqref{scheme} is a practical \emph{ansatz}, once the infinite dimensionality of Hilbert space has been recoded into a finite sum with bosonic-function coefficients. It is beyond the scope of this article to discuss it. We note only that it is more easily studied in Bargmann space~\cite{Sunko20,Sunko24}. There, the bijection between Hermite functions $\psi_n(x)$ in real space and complex numbers $u$ in Bargmann space, $\psi_n(x)\leftrightarrow u^n$, means that the $\Phi$'s have a natural cutoff, because high powers mean high-energy basis functions. The Bargmann-space picture is mathematically a complexification of the real-space formulation we adopt here, with the considerable advantage of a Hilbert-space structure in the complex numbers themselves. We choose the real-space approach because of its direct connection with already developed numerical implementations.

Let us specialize to fermions and $d=3$. There is a conjecture~\cite{Sunko20}, obviously extending to any odd $d$, that the shapes $S_i$ are precisely all the distinct iterated symmetrized derivatives $\prod_{a,b,c}\nabla^{(a,b,c)}\mathscr{D}_N$ of the triple product (the source shape)
\begin{equation}
\mathscr{D}_N = 
\widetilde{\Delta}_{x}\widetilde{\Delta}_{y}\widetilde{\Delta}_{z},
\label{triple}
\end{equation} 
where the symmetrized derivative is
\begin{equation}
\nabla^{(a,b,c)}=\sum_{i=1}^N\frac{\partial^a}{\partial x_i^a}
\frac{\partial^b}{\partial y_i^b}\frac{\partial^c}{\partial z_i^c}.
\end{equation} 
Here, $\widetilde{\Delta}_{x}$ denotes a normalized Vandermonde form in $N$ variables $x_1,\ldots,x_N$, defined in Eq.~\eqref{vdm} below, and $\widetilde{\Delta}_{y}$ and $\widetilde{\Delta}_{z}$ are the respective quantities for $y$ and $z$. From now on, we specialize to $N=3$, for which this conjecture is true. Derivatives of the Vandermonde form, known as \emph{harmonic polynomials}, read~\cite{Sunko22a}
\begin{equation}
x_{klm}=\begin{vmatrix}
\frac{x_{1\vphantom{g}}^{k}}{k!} & \frac{x_{2\vphantom{g}}^{l}}{l!} & \frac{x_{3\vphantom{g}}^{m}}{m!}\\
&&\\
\frac{x_{1\vphantom{g}}^{k-1}}{(k-1)!} & \frac{x_{2\vphantom{g}}^{l-1}}{(l-1)!} & \frac{x_{3\vphantom{g}}^{m-1}}{(m-1)!}\\
&&\\
\frac{x_{1\vphantom{g}}^{k-2}}{(k-2)!} & \frac{x_{2\vphantom{g}}^{l-2}}{(l-2)!} & \frac{x_{3\vphantom{g}}^{m-2}}{(m-2)!}
\end{vmatrix},\quad 0\le k,l,m\le 2,
\label{cvf}
\end{equation} 
where terms with negative factorials are zero. In particular,
\begin{equation}
x_{222} = \frac{1}{2}(x_1-x_2)(x_1-x_3)(x_2-x_3)
= \frac{1}{2}\Delta_{x} =: \widetilde{\Delta}_{x},
\label{vdm}
\end{equation} 
where $\Delta_{x}$ is the standard Vandermonde form in the three $x$-variables. Different permutations of $(k,l,m)$ may give rise to linearly independent polynomials. The number of independent harmonic polynomials of each degree $n$ is given by the coefficient of $q^n$ in the $q$-factorial~\cite{Humphreys90}, in this case:
\begin{equation}
[N]_q! = [3]_q! = 1\cdot (1+q)\cdot (1+q+q^2) = 1 + 2q + 2q^2 + q^3,
\end{equation} 
so there are six independent polynomials for $N=3$. They satisfy so-called \emph{syzygies},
\begin{equation}
\begin{split}
x_{112} + x_{121} + x_{211} &= 0, \\
x_{122} + x_{212} + x_{221} &= 0, \\
x_{121}x_{212} + x_{211}x_{212} + x_{121}x_{221} - 3x_{222} &= 0,
\end{split}
\label{syz}
\end{equation} 
so that $x_{112}$ and $x_{122}$ can be rewritten in terms of the others, which brings the number of independent first- and second-degree polynomials down to two, as predicted by the $q$-factorial. (Linear syzygies are often called rewriting rules.) The third (nonlinear) syzygy appears naturally in the evaluation of determinants presented below.

\subsection{The contemporary research context in combinatorics}\label{subsec4}

The cases with $d>1$ require $d$ sets of $N$ variables, one for each dimension of the laboratory space. They can be approached as an algorithmic problem: construct the generating set of all \emph{diagonally alternating} functions in $d$ sets of $N$ variables:
\begin{equation}
f(x_1,\ldots,z_N)=\mathrm{sign}(\sigma)f(x_{\sigma(1)},\ldots,z_{\sigma(N)}),
\label{alter}
\end{equation} 
where the permutation $\sigma$ acts diagonally on all $d$-plets, e.g., for $d=3$,
\begin{equation}
\sigma : (x_i,y_i,z_i) \mapsto (x_{\sigma(i)},y_{\sigma(i)},z_{\sigma(i)}), \quad i=1,\ldots,N.
\label{diag}
\end{equation} 
We denote the space of functions~\eqref{alter}, conforming to the alternating subgroup $\mathcal{A}_N$ of the diagonal action~\eqref{diag}, by $\mathscr{A}_N$, with $d=3$ unless stated otherwise. It can be presented as a free module. The generating set of this free module are the shapes, which are thus many-body invariants of the Pauli principle, the latter understood as invariance with respect to the symmetry group $\mathcal{A}_N$. For $d=1$, the generating set consists of the Vandermonde form alone, because it divides any antisymmetric polynomial in $N$ variables~\cite{Stanley99}. For $d=3$, as mentioned above, the generating set seems to comprise all the symmetrized derivatives $\prod_{a,b,c}\nabla^{(a,b,c)}\mathscr{D}_N$ of the form $\mathscr{D}_N$ in Eq.~\eqref{triple}~\cite{Sunko20}. This conjecture is contingent upon the choice of coefficient ring restricted to bosonic functions, as defined above.

The case $d=3$ has been studied previously with a different choice of coefficient ring~\cite{Bergeron13}, motivated by extending the $N!$ and $(N+1)^{N-1}$ hypotheses~\cite{Haiman03} from $d=2$ to $d=3$. The crucial difference with respect to the present approach is that, in the latter case, the coefficient ring consists of \emph{all} diagonally symmetric functions, which is larger than the ring of bosonic functions considered here. Variants of this problem are actively investigated in modern combinatorics, and can be technically demanding~\cite{Rhoades22}. Fortunately, our physical choice of coefficient ring simplifies matters somewhat, e.g., correspondences with classical results in $d=1$ are more transparent. Nevertheless, much is still unknown about the general case, so the present results are an important milestone in the further development of the physical theory.

\section{Main expressions}\label{main}

The main formal result of this work is a tool for extracting the content stipulated by Eq.~\eqref{scheme}. For an arbitrary wave function $\Psi\in\mathscr{A}_3$ in three vector variables $\mathbf{r}_i=(x_i,y_i,z_i)$, $i=1,2,3$, we derive a matrix $F$ such that
\begin{equation}
\Phi_i(v_0)=\sum_{j=0}^{35} F_{ij}\Psi(v_j),\quad i=0,\ldots,35,
\label{decomp}
\end{equation} 
where $v_j$ is a permutation of the nine variables $v_0=(x_1,\allowbreak x_2,\allowbreak x_3,\allowbreak y_1,\allowbreak y_2,\allowbreak y_3,\allowbreak z_1,\allowbreak z_2,\allowbreak z_3)$.

The first step is to limit the permutations in order to avoid the fermion sign problem~\cite{Hirsch85}. Namely, if $\Psi$ is diagonally alternating as described above, then any permutation of the triplets $\mathbf{r}_i$ gives rise to the same physical state. In order to remain in the same subspace \emph{modulo} such permutations, we allow only permutations of the $y_i$ and $z_i$ triplets in $v_0$:
\begin{equation}
v_i=\sigma_i v_0,
\end{equation} 
where $\sigma_i$ is an element of the direct product $\mathcal{S}_3\times \mathcal{S}_3=:\mathcal{S}_3^2$ of the symmetric group $\mathcal{S}_3$ with itself. It has $(3!)^2=36$ elements (listed in Table~\ref{varsSS3} in Appendix~\ref{conventab}), each an ordered pair of permutations acting independently on the two triplets. No information is lost by such a limitation, because three arbitrary permutations of the three triplets in $x$, $y$, $z$ can always be brought to a standard form by permuting the triplets $\mathbf{r}_1$, $\mathbf{r}_2$, $\mathbf{r}_3$ so that the $x$-triplet is brought to the identity permutation, e.g.,
\begin{equation}
(x_1, x_3, x_2, y_2, y_1, y_3, z_3, z_2, z_1)\to
(x_1, x_2, x_3, y_3, y_1, y_2, z_2, z_3, z_1),
\end{equation}
in this case, by exchanging particle indices $2$ and $3$.

Notably, $\mathcal{S}_3^2$ has the same number of elements as $\mathscr{A}_3$ has generators as a module. Fixing one set of $N$ variables in $d$ dimensions leaves the group $\mathcal{S}_N^{d-1}$ to act on the rest, and the number of generators of the free module is also $(N!)^{d-1}$ in general. Therefore, the abstract presentation of $\mathscr{A}_N$ in $d$ dimensions as a module generated by shapes can always be coordinatized by the action of $\mathcal{S}_N^{d-1}$ on a specific $\Psi\in\mathscr{A}_N$, i.e., by evaluating $\Psi(\sigma v_0)$ for all $\sigma\in\mathcal{S}_N^{d-1}$. We implement this program for $(N,d)=(3,3)$ now, introducing the technique with two simple examples of increasing complexity.

\subsection{Two fermions in three dimensions}\label{twothree}

The case $N=2$ is the simplest example of this scheme. We have
\begin{equation}
\mathscr{D}_2=\begin{vmatrix}
x_1 & x_2 \\
1 & 1 
\end{vmatrix}
\begin{vmatrix}
y_1 & y_2 \\
1 & 1 
\end{vmatrix}
\begin{vmatrix}
z_1 & z_2 \\
1 & 1 
\end{vmatrix}
=(x_1-x_2)(y_1-y_2)(z_1-z_2).
\end{equation} 
The non-zero derivatives are
\begin{equation}
\nabla^{(1,1,0)}\mathscr{D}_2 = 2(z_1-z_2),
\end{equation} 
and cyclically, accounting for the four shapes, which could have been written down by inspection. Denoting $x=x_1-x_2$ etc., any $\Psi\in\mathscr{A}_2$ can be written
\begin{equation}
\Psi(x_1,x_2,y_1,y_2,z_1,z_2) = \Phi_0xyz + \Phi_1x + \Phi_2y + \Phi_3z,
\label{2part}
\end{equation} 
where the $\Phi_i$ are symmetric under the transpositions $x_1\leftrightarrow x_2$ etc.\ in each spatial direction separately. [Thus, $xyz$ is an independent generator despite (say) $xy$ being a symmetric function, because $x$ and $y$ are antisymmetric.] This expression can be evaluated at three such transpositions to yield
\begin{equation}
\begin{pmatrix*}[l]
\Psi \\ \Psi_x \\ \Psi_y \\ \Psi_z
\end{pmatrix*}
=
\begin{pmatrix*}[r]
xyz & x & y & z \\
-xyz & -x & y & z \\
-xyz & x & -y & z \\
-xyz & x & y & -z
\end{pmatrix*}
\begin{pmatrix}
\Phi_0 \\ \Phi_1 \\ \Phi_2 \\ \Phi_3
\end{pmatrix},
\label{trans2}
\end{equation} 
where the indices in the left-hand column indicate which pair of variables is transposed. The matrix in Eq.~\eqref{trans2} is easily inverted to yield the bosonic functions $\Phi_0, \ldots, \Phi_3$, as long as $xyz\neq 0$. The reader will recognize a generalization of the well-known separation of a function into symmetric and antisymmetric parts. In the following, the generalization becomes more complex, but the idea remains the same.

The compactness of the shape formulation becomes apparent upon its direct comparison with Slater determinants. For example, the polynomial part of a simple doubly-excited state may be written $(x^2+y^2+z^2)x$. In Slater determinants, this expression reads
\begin{equation}
\begin{split}
(x^2+y^2+z^2)x=&\begin{vmatrix}x_1^3 & x_2^3\\ 1 & 1\end{vmatrix}
+\begin{vmatrix}x_1^2 & x_2^2\\ x_1 & x_2\end{vmatrix}
+\begin{vmatrix}x_1y_1^2 & x_2y_2^2\\ 1 & 1\end{vmatrix}
-\begin{vmatrix}y_1^2 & y_2^2\\ x_1 & x_2\end{vmatrix} \\
&+2\begin{vmatrix}x_1y_1 & x_2y_2\\ y_1 & y_2\end{vmatrix}
+\begin{vmatrix}x_1z_1^2 & x_2z_2^2\\ 1 & 1\end{vmatrix}
-\begin{vmatrix}z_1^2 & z_2^2\\ x_1 & x_2\end{vmatrix}
+2\begin{vmatrix}x_1z_1 & x_2z_2\\ z_1 & z_2\end{vmatrix},
\end{split}
\end{equation}
which appears rather unwieldly. For a complete wave function, replace each entry $x_i^ly_j^mz_k^n$ with the Hermite functions $\psi_l(x_i)\psi_m(y_j)\psi_n(z_k)$, with missing factors as zero powers.

\subsection{Three particles in one dimension}\label{threeone}

In this example, we allow the function $\Psi$ to be of any symmetry. Then the role of shapes is played by all harmonic polynomials. We have
\begin{equation}
\Psi(x_1,x_2,x_3) = \Phi_0x_{222}+\Phi_1x_{212}+\Phi_2x_{221}
 + \Phi_3x_{211}+\Phi_4x_{121} + \Phi_5x_{210},
\end{equation} 
where the $\Phi_i$ are symmetric, i.e., constant under the action of $\mathcal{S}_3$, the group of permutations of the three variables. Note that $x_{210}=-1$, so we shall keep the ``210'' polynomials only for uniformity. We need a generalization of the separation into symmetric and antisymmetric parts. One such generalization is the Fourier transform, where ``$+1$'' and ``$-1$'' generalize to complex roots of unity. It is appropriate for periodic functions, which are invariant under the cyclic group, whose characters are complex roots of unity. Here, the required transform is by the characters $\chi_R(\sigma)$ of $\mathcal{S}_3$.  The character table is~\cite{Meliot17}
\begin{equation}
\begin{array}{cc|cccccc}
R & \sigma: & (1)(2)(3) & (23)(1) & (12)(3) & (13)(2) & (123) & (132) \\
\hline
S & & \phantom{-}1 & \phantom{-}1 & \phantom{-}1 & \phantom{-}1 & \phantom{-}1 & \phantom{-}1 \\
A & & \phantom{-}1 & -1 & -1 & -1 & \phantom{-}1 & \phantom{-}1 \\
E & & \phantom{-}2 & \phantom{-}0 & \phantom{-}0 & \phantom{-}0 & -1 & -1
\end{array},
\label{S3chars}
\end{equation} 
the rows referring respectively to the symmetric, antisymmetric, and two-dimensional representations $R$. The columns list the corresponding six permutations in cycle notation, e.g., $(12)(3)$ is the transposition $x_1\leftrightarrow x_2$. We write down the transforms
\begin{equation}
g_R(x_1,x_2,x_3) = \sum_\sigma \chi_R(\sigma)
\Psi(x_{\sigma(1)},x_{\sigma(2)},x_{\sigma(3)})
\end{equation} 
explicitly:
\begin{equation}
\begin{aligned}
g_S & = 6\Phi_5x_{210} = -6\Phi_5,\\
g_A & = 6\Phi_0x_{222} = 3\Phi_0\Delta_x,\\
g_E & = 3(\Phi_1x_{212}+\Phi_2x_{221}
 + \Phi_3x_{211}+\Phi_4x_{121}).
\end{aligned}
\end{equation} 
The character $\chi_E$ extracts all four harmonic polynomials of mixed symmetry. In order to obtain the bosonic functions $\Phi_1,\ldots, \Phi_4$, we need to evaluate $g_E$ at additional permuted sets of variables. Writing $g_{klm}$ for $g_E(x_k,x_l,x_m)$, we have
\begin{equation}
\begin{pmatrix*}[l]
g_{123}\\g_{132}\\g_{213}\\g_{312}
\end{pmatrix*}
=
3\begin{pmatrix}
x_{212} & x_{221} & x_{211} & x_{121} \\
-x_{221} & -x_{212} & -x_{211} & x_{211} + x_{121} \\
x_{212}+x_{221} & -x_{221} & -x_{121} & -x_{211} \\
-x_{212}-x_{221} & x_{212} & -x_{211}-x_{121} & x_{211} \\
\end{pmatrix}
\begin{pmatrix*}[l]
\Phi_1\\\Phi_2\\\Phi_3\\\Phi_4
\end{pmatrix*}.
\label{ctog}
\end{equation} 
Importantly, the determinant of the matrix in Eq.~\eqref{ctog} is just the discriminant $\Delta_x^2$, up to a constant factor, so there are no additional requirements for invertibility other than that none of the points $x_i$ coincide. The rank of the matrix depends on the choice of variable permutations at which $g_E$ is evaluated, which we found by trial and error here. It does not matter which set of permuted variables is chosen, as long as it gives a full-rank matrix without additional conditions on the variables.

\subsection{Three fermions in three dimensions}\label{threethree}

Now, we are set to derive the main result. Because the expressions turn out to be quite large, we relegate most of them to Appendix~\ref{matrixel}, tracing only the main steps here. The technical approach employed is to express both the shapes and the matrix elements of the transformations like~\eqref{ctog} in terms of the harmonic polynomials, treated as new variables, and to apply the syzygies~\eqref{syz} to simplify the resulting expressions when necessary.

Write an arbitrary wave function as
\begin{equation}
\Psi(v_0) = \sum_{i=0}^{35} \Phi_i(v_0)S_i(v_0),
\label{Psiv0}
\end{equation} 
where $\Phi_i$ are functions symmetric in each direction in space separately, so that
\begin{equation}
\Phi_j(\sigma_iv_0)=\Phi_j(v_0)=:\Phi_j\quad 0\le i,j \le 35.
\end{equation} 
The quantities $S_i$, $i = 0, \ldots, 35$ are the shapes compiled in Table~\ref{evalS} in Appendix~\ref{conventab} (with $S_0=\mathscr{D}_3$). The character table of $\mathcal{S}_3^2$ (listed in Table~\ref{varcharsSS3}) is the outer product of the character table~\eqref{S3chars} with itself. The ensuing $9$ characters are evaluated at the $36$ permutations $\sigma_i\in \mathcal{S}_3^2$. We analyze the wave function~\eqref{Psiv0} with these characters, calling the respective transforms $g_i$:
\begin{equation}
g_i(v_0) := \sum_{\sigma}\chi_i(\sigma)\Psi(\sigma v_0).
\label{chitransform}
\end{equation} 
The first four characters each extract one term from Eq.~\eqref{Psiv0}:
\begin{equation}
\begin{split}
g_0(v_0) &= 36 \Phi_{32}S_{32} = 216 \Phi_{32}x_{222},\\
g_1(v_0) &= 36 \Phi_{0}S_{0} = 36\Phi_0x_{222}y_{222}z_{222},\\
g_2(v_0) &= 36 \Phi_{23}S_{23} = 216 \Phi_{23}z_{222},\\
g_3(v_0) &= 36 \Phi_{26}S_{26} = 216 \Phi_{26}y_{222}.
\end{split}
\label{g0123}
\end{equation} 
Thus, we obtain the bosonic functions $\Phi_{32}$, $\Phi_{0}$, $\Phi_{23}$, and $\Phi_{26}$ simply by dividing these $g_i$'s with some Vandermonde forms. The manifest asymmetry of these expressions is due to our choice to keep the $x$-triplet fixed in the action of $\mathcal{S}_3^2$, so the trivial (identity) character extracts $x_{222}$.

The next four characters extract four terms each:
\begin{equation}
\begin{split}
g_4(v_0) &= 18 \left(\Phi_{1}S_{1} + \Phi_{4}S_{4} + \Phi_{6}S_{6} + \Phi_{11}S_{11}\right),\\
g_5(v_0) &= 18 \left(\Phi_{22}S_{22} + \Phi_{29}S_{29} + \Phi_{31}S_{31} + \Phi_{35}S_{35}\right),\\
g_6(v_0) &= 18 \left(\Phi_{2}S_{2} + \Phi_{5}S_{5} + \Phi_{8}S_{8} + \Phi_{13}S_{13}\right),\\
g_7(v_0) &= 18 \left(\Phi_{20}S_{20} + \Phi_{27}S_{27} + \Phi_{30}S_{30} + \Phi_{34}S_{34}\right).
\end{split}
\end{equation} 
Like in the reduced example in Section~\ref{threeone}, we can obtain the $\Phi_i$'s by evaluating these expressions at other sets of variables in order to construct $4\times 4$ systems of equations for them. For $g_4$, we find that the variable set $\{v_0, v_6, v_7, v_{24}\}$ gives a full-rank matrix in the equation
\begin{equation}
\begin{pmatrix*}[l]
g_4(v_0)\\g_4(v_6)\\g_4(v_7)\\g_4(v_{24})
\end{pmatrix*}
=18
\begin{pmatrix*}[l]
S_1(v_0) & S_4(v_0) & S_6(v_0) & S_{11}(v_0) \\
S_1(v_6) & S_4(v_6) & S_6(v_6) & S_{11}(v_6) \\
S_1(v_7) & S_4(v_7) & S_6(v_7) & S_{11}(v_7) \\
S_1(v_{24}) & S_4(v_{24}) & S_6(v_{24}) & S_{11}(v_{24}) \\
\end{pmatrix*}
\begin{pmatrix*}[l]
\Phi_1\\\Phi_4\\\Phi_6\\\Phi_{11}
\end{pmatrix*},
\label{ctog4}
\end{equation} 
which is analogous to~\eqref{ctog}. The elements of the matrix $M_4$ that appears in the inverse equation,
\begin{equation}
\begin{pmatrix*}[l]
\Phi_1\\\Phi_4\\\Phi_6\\\Phi_{11}
\end{pmatrix*}
=\frac{2^2}{3^5}
\frac{1}{\Delta_{x}\Delta_{y}\Delta_{z}}
M_4\begin{pmatrix*}[l]
g_4(v_0)\\g_4(v_6)\\g_4(v_7)\\g_4(v_{24})
\end{pmatrix*},
\label{gtoc4}
\end{equation} 
are listed in Table~\ref{tableM4}. Notably, $M_4$ is $z$-independent, but the particular transforms $g_4(v_j)$ appearing in Eq.~\eqref{gtoc4} are all divisible by $\Delta_z$, so the extra $\Delta_z$ in~\eqref{gtoc4} is cancelled. Finally, it is recovered in the expansion because the shapes $S_1$, $S_4$, $S_6$, and $S_{11}$ all contain a factor $z_{222}$. Hence, the character $\chi_4$ extracts components of the wave function symmetric in the $xy$ plane, multiplied by a term $\Delta_z$ which is fully antisymmetric in the $z$ direction.

\begin{table}
\begin{center}
\begin{tabular}{|c|c|c|c|}
\hline
$(i,j)$ & $M_4(i,j)$ & $(i,j)$ & $M_4(i,j)$ \\ \hline
$(1,1)$ & $x_{121}y_{121} + x_{211}y_{121} + x_{211}y_{211}$ & $(3,1)$ & $x_{121}y_{212} - x_{211}y_{221}$ \\
$(1,2)$ & $-x_{121}y_{121} + x_{211}y_{211}$ & $(3,2)$ & $-x_{121}y_{212} - x_{211}y_{212} - x_{211}y_{221}$ \\
$(1,3)$ & $x_{211}y_{121} + x_{121}y_{211} + x_{211}y_{211}$ & $(3,3)$ & $-x_{121}y_{212} - x_{121}y_{221} - x_{211}y_{221}$ \\
$(1,4)$ & $x_{211}y_{121} - x_{121}y_{211}$ & $(3,4)$ & $x_{121}y_{212} + x_{211}y_{212} + x_{121}y_{221}$ \\
$(2,1)$ & $-x_{221}y_{121} - x_{212}y_{211} - x_{221}y_{211}$ & $(4,1)$ & $x_{212}y_{212} + x_{212}y_{221} + x_{221}y_{221}$ \\
$(2,2)$ & $-x_{212}y_{121} - x_{212}y_{211} - x_{221}y_{211}$ & $(4,2)$ & $x_{221}y_{212} + x_{212}y_{221} + x_{221}y_{221}$ \\
$(2,3)$ & $-x_{212}y_{121} - x_{221}y_{121} - x_{221}y_{211}$ & $(4,3)$ & $-x_{212}y_{212} + x_{221}y_{221}$ \\
$(2,4)$ & $-x_{212}y_{121} - x_{221}y_{121} - x_{212}y_{211}$ & $(4,4)$ & $-x_{221}y_{212} + x_{212}y_{221}$ \\ \hline
\end{tabular}
\end{center}
\caption{Matrix elements of the matrix $M_4$ in Eq.~\eqref{gtoc4}.}
\label{tableM4}
\end{table}

Similarly, for $g_5$ we obtain
\begin{equation}
\begin{pmatrix*}[l]
\Phi_{22}\\\Phi_{29}\\\Phi_{31}\\\Phi_{35}
\end{pmatrix*}
=\frac{2}{3^6}
\frac{1}{\Delta_{x}\Delta_{y}}
M_5\begin{pmatrix*}[l]
g_5(v_0)\\g_5(v_6)\\g_5(v_7)\\g_5(v_{24})
\end{pmatrix*},
\label{gtoc5}
\end{equation} 
where the matrix $M_5$ is given in Appendix~\ref{matrixel}. It is also $z$-independent, but it yields a different structure. The transforms $g_5(v_j)$ are symmetric in the $z$ direction, and the shapes $S_{22}$, $S_{29}$, $S_{31}$, and $S_{35}$ all contain $z_{210}=-1$, so they are $z$-independent. Hence, the character $\chi_5$ extracts a two-dimensional component of $\Psi$ that is antisymmetric in the $xy$-plane, multiplied by a symmetric $z$-component that is simply copied over to the corresponding bosonic functions $\Phi_i$.

The next two characters repeat this pattern in the $y$ direction. We have
\begin{equation}
\begin{pmatrix*}[l]
\Phi_{2}\\\Phi_{5}\\\Phi_{8}\\\Phi_{13}
\end{pmatrix*}
=\frac{2^2}{3^5}
\frac{1}{\Delta_{x}\Delta_{y}\Delta_{z}}
M_6\begin{pmatrix*}[l]
g_6(v_0)\\g_6(v_1)\\g_6(v_2)\\g_6(v_{4})
\end{pmatrix*}
\label{gtoc6}
\end{equation} 
and
\begin{equation}
\begin{pmatrix*}[l]
\Phi_{20}\\\Phi_{27}\\\Phi_{30}\\\Phi_{34}
\end{pmatrix*}
=\frac{2}{3^6}
\frac{1}{\Delta_{x}\Delta_{z}}
M_7\begin{pmatrix*}[l]
g_7(v_0)\\g_7(v_1)\\g_7(v_2)\\g_7(v_{4})
\end{pmatrix*},
\label{gtoc7}
\end{equation} 
where the matrices $M_6$ and $M_7$, given in Appendix~\ref{matrixel}, are both $y$-independent. The transforms $g_6$ all contain a Vandermonde factor $\Delta_y$ cancelled by the denominator in Eq.~\eqref{gtoc6}, while the shapes $S_{2}$, $S_{5}$, $S_{8}$, and $S_{13}$ are proportional to $y_{222}$. The transforms $g_7$ are symmetric in the $y$ direction, while the shapes $S_{20}$, $S_{27}$, $S_{30}$, and $S_{34}$ all contain $y_{210}=-1$.

\begin{table}
\begin{equation*}
Gg = \frac{1}{3}\begin{pmatrix}
 [(v_{4}) - (v_{10}) + (v_{12}) - (v_{24})]/2 \\
 [(v_{0}) + (v_{9}) + (v_{10}) + (v_{14}) + (v_{24}) + (v_{32})]/2 \\
 [(v_{1}) + (v_{2}) + (v_{6}) + (v_{7}) + (v_{18}) + (v_{26})]/2 \\
 [(v_{2}) - (v_{6}) + (v_{23}) + (v_{26}) + (v_{28})]/2 \\ \hline
 [-(v_{4}) - (v_{10}) + (v_{12}) + (v_{24})]/2 \\
 [(v_{1}) + (v_{2}) - (v_{6}) - (v_{7}) - (v_{18}) + (v_{26})]/2 \\
 [-(v_{1}) + 2(v_{6}) + (v_{7}) + (v_{18}) - (v_{23}) + (v_{28})]/2 \\
 [(v_{0}) + (v_{4}) - (v_{9}) - (v_{12}) - (v_{14}) + (v_{32})]/2 \\ \hline
 (v_{0}) - (v_{9}) - (v_{10}) - (v_{12}) - (v_{32}) \\
 (v_{1}) - (v_{7}) + (v_{18}) + (v_{26}) + (v_{28}) \\
 (v_{2}) + (v_{7}) - (v_{23}) - (v_{26}) \\
 (v_{4}) + (v_{9}) - (v_{14}) + (v_{24}) + (v_{32}) \\
 -(v_{1}) + (v_{2}) - (v_{7}) - (v_{18}) + 2(v_{23}) - 2(v_{26}) - (v_{28}) \\
 2(v_{1}) + (v_{2}) + 2(v_{7}) - (v_{18}) - (v_{23}) + (v_{26}) + 2(v_{28}) \\
-2(v_{0}) - (v_{4}) + (v_{9}) - (v_{10}) - (v_{12}) - 2(v_{14}) - (v_{24}) + (v_{32}) \\
 (v_{0}) - (v_{4}) + (v_{9}) + 2(v_{10}) + 2(v_{12}) + (v_{14}) - (v_{24}) - 2(v_{32})
 \end{pmatrix}
\end{equation*}
\caption{The column $Gg$ in Eq.~\eqref{gtoc8}. The blocks in Eq.~\eqref{Utildeinv} are separated by horizontal lines.}
\label{tableGg}
\end{table} 

For $\chi_8$, the situation is more complicated. It extracts $16$ terms:
\begin{equation}
\begin{split}
g_8(v_0) & = 9\left(\Phi_{3}S_{3} + \Phi_{7}S_{7} + \Phi_{9}S_{9} + \Phi_{10}S_{10} + \Phi_{12}S_{12} + \Phi_{14}S_{14}\, + \right. \\
&\phantom{9[\Phi_{3}S_{3}} + \Phi_{15}S_{15} + \Phi_{16}S_{16} + \Phi_{17}S_{17} + \Phi_{18}S_{18} + \Phi_{19}S_{19}\, + \\
&\left.\phantom{9[\Phi_{3}S_{3}} + \Phi_{21}S_{21} + \Phi_{24}S_{24} + \Phi_{25}S_{25} + \Phi_{28}S_{28} + \Phi_{33}S_{33}\right).
\end{split}
\end{equation} 
Evaluating it at $16$ particular sets of variables, one arrives at the rank-$16$ matrix $U$ in $g=U\Phi$, as in Eq.~\eqref{ctog4}. Next, we use Gaussian (row) elimination $G$ and column reorderings $C$ to bring it into block-matrix form $\widetilde{U} = GUC$:
\begin{equation}
g=G^{-1}\widetilde{U}C^{-1}\Phi\implies C^{-1}\Phi = \widetilde{U}^{-1} Gg.
\label{gtoc8}
\end{equation} 
The matrix $C^{-1}$ acting on $\Phi$ from the left just reorders the $\Phi_i$'s, while the Gaussian-elimination matrix $G$ induces linear relations among the $g_8(v_i)$'s. Thus,
\begin{equation}
\begin{split}
C^{-1}\Phi = \mathrm{col}\, 
[\,& \Phi_{3} , \Phi_{9}, \Phi_{10}, \Phi_{16},|\, \Phi_{17}, \Phi_{24}, \Phi_{25}, \Phi_{33},
\\
& |\,\Phi_{7}, \Phi_{12}, \Phi_{14}, \Phi_{15}, \Phi_{18}, \Phi_{19}, \Phi_{21}, \Phi_{28}\,]
\end{split}
\end{equation} 
where the additional vertical lines indicate the blocking of $\widetilde{U}^{-1}$ into two $4\times 4$ and one $8\times 8$ block. Writing $(v_k)$ for $g_8(v_k)$, the column $Gg$ is given in Table~\ref{tableGg}. The matrix $\widetilde{U}^{-1}$ is
\begin{equation}
\widetilde{U}^{-1}=\frac{2^3}{3^5}\frac{1}{\Delta_{x}\Delta_{y}\Delta_{z}}
\begin{pmatrix}
M_8 & \emptyset & \emptyset \\
\emptyset & \frac{\Delta_{x}}{2} M_9 & \emptyset \\
\emptyset & \emptyset & \frac{1}{3} M_{10}
\end{pmatrix}.
\label{Utildeinv}
\end{equation} 
The $x$-independent $4\times 4$ matrices $M_8$ and $M_9$ repeat the pattern of $M_4$ and $M_5$ (or $M_6$ and $M_7$), respectively. The shapes $S_{3}$, $S_{9}$, $S_{10}$, and $S_{16}$ are proportional to $\Delta_x$, while $S_{17}$, $S_{24}$, $S_{25}$, and $S_{33}$ all contain $x_{210}=-1$. Again, the matrices $M_8$, $M_9$, and $M_{10}$ are given in Appendix~\ref{matrixel}. No special meaning should be ascribed to the appearance of $M_8$ and $M_9$ only after the reduction of $\chi_8$, while the four matrices referring to the other two directions each appeared with a character of its own. That is again due to the asymmetric action of $\mathcal{S}_3^2$, where the $x$-triplet is kept constant. Similarly, the extra work involved in the column $Gg$ simply recovers the symmetry of the bosonic functions in the $x,y,z$ directions that was broken by this choice.

The characters of $\mathcal{S}_3^2$ break up the spatial dependence of an arbitrary function $\Psi\in\mathscr{A}_3$ into eleven blocks. Four are one-dimensional, comprising the product $x_{222}y_{222}z_{222}$ and its factors $x_{222}$, $y_{222}$, and $z_{222}$. Three are four-dimensional, with a symmetric spatial dependence in a plane, and antisymmetric in the line orthogonal to that plane. Another three are also four-dimensional, antisymmetric in a plane and symmetric in the orthogonal direction. Finally, there is one eight-dimensional block whose antisymmetry cannot be decomposed across the lower spatial dimensions. The total decomposition reads
\begin{equation}
1+ 3\times 1 + 3\times 4 +3\times 4 + 8 = 36.
\end{equation} 
The question arises whether the seven higher-dimensional blocks are reducible. That is the subject of the next section.

The program to extract the bosonic functions $\Phi_i$ by evaluations of an arbitrary wave function $\Psi\in\mathscr{A}_3$ is now complete. To summarize: Evaluate the transforms~\eqref{chitransform} at the required sets of variables, insert them in the right-hand columns of the matrix equations~\eqref{gtoc4}, \eqref{gtoc5}, \eqref{gtoc6}, \eqref{gtoc7}, and \eqref{gtoc8}, and multiply by the matrices $M_i$ as given above. Thus, one obtains the matrix $F_{ij}$ in the formula~\eqref{decomp} for the bosonic functions $\Phi_i$, $i=0, \ldots, 35$.

\subsection{The symmetry of the coefficient ring}

\begin{table}
\renewcommand*{\arraystretch}{1.2}
\begin{center}
\begin{tabular}{|rr|rr|rr|rr|rr|rr|rr|rrr|}
\hline
$I_0$ & 32 & $I_4$ &  1 & $I_5$ & 22 & $I_6$ &  2 & $I_7$ & 20 & $I_8$ &  3 & $I_9$ & 17 & $I_{10}$ &  7 & 18 \\ \cline{1-2}
$I_1$ &  0 &       &  4 &       & 29 &       &  5 &       & 27 &       &  9 &       & 24 &       & 12 & 19 \\ \cline{1-2}
$I_2$ & 23 &       &  6 &       & 31 &       &  8 &       & 30 &       & 10 &       & 25 &       & 14 & 21 \\ \cline{1-2}
$I_3$ & 26 &       & 11 &       & 35 &       & 13 &       & 34 &       & 16 &       & 33 &       & 15 & 28 \\ \hline
\end{tabular}
\end{center}
\caption{Indices of shapes appearing in the blocks in the inversion~\eqref{schemeinv}.
\label{blocktable}}
\end{table}

Once the steps mentioned in the previous paragraph are carried out, one can insert the expression~\eqref{decomp} for the bosonic functions into the decomposition~\eqref{scheme} to obtain the inversion formula:
\begin{equation}
\begin{split}
\Psi(v_0) & = \sum_{i=0}^{35}\Phi_iS_i=\sum_{k=0}^{10}\sum_{i\in I_k}\Phi_iS_i \\
& = \sum_{k=0}^{10}\sum_{i\in I_k} \sum_{j=0}^{35} S_i F_{ij} \Psi(v_j)
=:\sum_{j=0}^{35}
\left[\frac{1}{36}\sum_{k=0}^{10}\bar{\eta}_k(\sigma_j)\right]\Psi(\sigma_j v_0),
\end{split}
\label{schemeinv}
\end{equation} 
where the sets $I_k$ of shape indices that appear in the blocks above (termed shape blocks in the following) are listed in Table~\ref{blocktable}, and the newly introduced characters read
\begin{equation}
\bar{\eta}_k(\sigma_j) = 36\sum_{i\in I_k} S_iF_{ij},
\quad k=0,\ldots,10;\; j=0,\ldots,35.
\end{equation} 
These characters, listed in Table~\ref{deltamat} in the Appendix~\ref{conventab}, are just numbers, confirming that the inversion formula~\eqref{schemeinv} is well defined. They add column-wise to the Kronecker delta,
\begin{equation}
\frac{1}{36}\sum_{k=0}^{10}\bar{\eta}_k(\sigma_j)=\delta_{j0},\quad j=0,\ldots, 35,
\end{equation} 
in accord with the inversion~\eqref{schemeinv}. Summing their squares row-wise yields
\begin{equation}
\frac{1}{36}\sum_{j=0}^{35}\bar{\eta}_k(\sigma_j)^2 = |I_k|,
\quad k=0,\ldots, 10,
\end{equation} 
where $|I_k|$ is the number of elements in the set $I_k$, which means that for $k\ge 4$, each $\bar{\eta}_k$ is a composite character with $|I_k|$ components. The latter are one-dimensional (symmetric) representations of the symmetry group $\mathcal{S}_3^2$ of the coefficient ring, which coordinatizes it in the same way as $\mathscr{A}_3$ (keeping one triplet of variables fixed). Since, intuitively, they appear when $F_{ij}$ ``looks to the left'' in the inversion formula~\eqref{schemeinv}, they are dual to the bosonic functions $\Phi_i$. They are symmetric functions, because the alternating properties of the shapes are compensated by the Vandermonde denominators in $F_{ij}$. There are $36$ of them because there are that many ways to pass from the symmetry of each individual shape to the symmetry of the coefficient ring. We shall not pursue this reduction further here, because it distracts from our main purpose of applying the invariant (isotypical) qualitative analysis of the wave function, already accomplished by its decomposition into the aforementioned $11$ shape blocks.

\section{A numerical example: The lowest-energy $^4P_u$ electronic state of the lithium atom}\label{sec5}

The ideas expounded above are vividly illustrated by the lowest-energy  $^4 P_u$ quartet state of the lithium atom, whose nonrelativistic Born-Oppenheimer description involves an electronic wave function given by a product of a spin component and its spatial counterpart $\Psi := \Psi(\mathbf{r}_1,\mathbf{r}_2,\mathbf{r}_3)$ that is approximately given by the Slater determinant built from the $1s$, $2s$, and $2p_z$ orbitals. Electron correlation effects beyond this zeroth-order approximation are quite weak, as the permutational antisymmetry of $\Psi$ reduces the electron-electron repulsion considerably. They are efficiently accounted for with the \emph{ansatz}
\begin{equation}
\begin{split}
\Psi\approx  \hat A \sum_{n=1}^\mathcal{N} C_n & \exp \big( 
\alpha_{1n} |\mathbf{r}_1|^2 + \alpha_{2n} |\mathbf{r}_2|^2 + \alpha_{3n} |\mathbf{r}_3|^2 + \beta_{1n} |\mathbf{r}_1-\mathbf{r}_2|^2
\\ &
+ \beta_{2n} |\mathbf{r}_1-\mathbf{r}_3|^2+ \beta_{3n} |\mathbf{r}_2-\mathbf{r}_3|^2 \big) \sinh (\gamma_{1n} z_1+\gamma_{2n} z_2+\gamma_{3n} z_3) ,
\end{split}
\label{a1}
\end{equation} 
where $ \hat A$ is the antisymmetrizer. The coefficients $\{ C_n \}$ and the nonlinear parameters  $\{ \alpha_{1n},  \alpha_{2n},  \alpha_{3n},  \beta_{1n}, \beta_{2n},  \beta_{3n}, \gamma_{1n}, \gamma_{2n},  \gamma_{3n} \}$ that parametrize the $\mathcal{N}$ explicitly correlated Gaussians (ECGs) ~\cite{Mitroy13,Bubin13} are determined by minimizing the expectation value of the electronic energy $\langle \Psi | \hat H | \Psi \rangle$, where
\begin{equation}
\hat H =\sum_{i=1}^3 \left( \hspace{-3pt} -\frac{1}{2}  \mathbf{\nabla}_i^2 - 3   |\mathbf{r}_i|^{-1} \right) 
+\sum_{i>j=1}^3 |\mathbf{r}_i-\mathbf{r}_j|^{-1},
\label{a2}
\end{equation} 
subject to the normalization condition 
$\langle \Psi | \Psi \rangle = 1$ (atomic units are used here and in the following).  The minimization involves the usual secular-equation formalism for the coefficients and various numerical techniques for the nonlinear parameters ~\cite{Mitroy13,Bubin13}.  The ECG basis sets are constructed iteratively from an initial set of three functions with $\mathcal{\mathcal{N}}_1=1$, $\mathcal{N}_2=2$, and $\mathcal{N}_3=3$.  For each $J > 3$, the sets of $\mathcal{N}_{J-2}$ and $\mathcal{N}_{J}$ ECGs with the nonlinear parameters already optimized by separate minimizations are merged and the resulting parameters of the set of $\mathcal{N}_{J+1} =\mathcal{N}_{J-2}+\mathcal{N}_{J}$ ECGs are employed as the initial guess for the next-stage optimization.\footnote{The numbers $\mathcal{N}_1$, $\mathcal{N}_2$, \ldots, are the elements of the so-called Narayana's cows sequence; see The Online Encyclopedia of Integer Sequences, https://oeis.org/A000930.}  The computed approximate wave functions and the corresponding energies converge rapidly with $\mathcal{N}$ (Table~\ref{tabconv}).  Although, strictly speaking, the wave functions constructed from the lobe ECGs that enter Eq.~(\ref{a1}) are not eigenfunctions of the total angular momentum operator $\hat L^2$, the deviations are numerically insignificant.

\addtolength{\tabcolsep}{-4pt}
\begin{table}
{\scriptsize
\renewcommand*{\arraystretch}{1.3}
\begin{tabular}{|c|c|c|c|c|c|c|c|c|}
\hline
$\mathcal{N}$ & $9$ & $13$ & $19$ & $28$ & $41$ & $129$ & $277$ & $1278$ \\
\hline
$\langle \Psi|\hat H | \Psi\rangle$ & $-5.3580443$ & $-5.3631457$ & $-5.3661042$ & $-5.3672326$ & $-5.3676610$ & $-5.3679944$ & $-5.3680085$ & $-5.3680101$ \\
   $w_0$ & $9.4913\cdot 10^{-4}$ & $9.1663\cdot 10^{-4}$ & $9.3229\cdot 10^{-4}$ & $9.0504\cdot 10^{-4}$ & $9.0971\cdot 10^{-4}$ & $9.1230\cdot 10^{-4}$ & $9.1259\cdot 10^{-4}$ & $9.1257\cdot 10^{-4}$ \\
   $w_1$ & $4.2089\cdot 10^{-5}$ & $3.5268\cdot 10^{-5}$ & $3.6106\cdot 10^{-5}$ & $3.4150\cdot 10^{-5}$ & $3.4398\cdot 10^{-5}$ & $3.3406\cdot 10^{-5}$ & $3.3411\cdot 10^{-5}$ & $3.3414\cdot 10^{-5}$ \\
   $w_2$ & $2.6900\cdot 10^{-1}$ & $2.6987\cdot 10^{-1}$ & $2.7084\cdot 10^{-1}$ & $2.7174\cdot 10^{-1}$ & $2.7205\cdot 10^{-1}$ & $2.7208\cdot 10^{-1}$ & $2.7209\cdot 10^{-1}$ & $2.7208\cdot 10^{-1}$ \\
   $w_3$ & $9.4913\cdot 10^{-4}$ & $9.1663\cdot 10^{-4}$ & $9.3229\cdot 10^{-4}$ & $9.0504\cdot 10^{-4}$ & $9.0971\cdot 10^{-4}$ & $9.1230\cdot 10^{-4}$ & $9.1259\cdot 10^{-4}$ & $9.1257\cdot 10^{-4}$ \\
   $w_4$ & $5.1469\cdot 10^{-2}$ & $5.1525\cdot 10^{-2}$ & $5.1492\cdot 10^{-2}$ & $5.1512\cdot 10^{-2}$ & $5.1549\cdot 10^{-2}$ & $5.1493\cdot 10^{-2}$ & $5.1494\cdot 10^{-2}$ & $5.1493\cdot 10^{-2}$ \\
   $w_5$ & $3.2297\cdot 10^{-3}$ & $3.1283\cdot 10^{-3}$ & $3.1829\cdot 10^{-3}$ & $3.1083\cdot 10^{-3}$ & $3.1234\cdot 10^{-3}$ & $3.1346\cdot 10^{-3}$ & $3.1352\cdot 10^{-3}$ & $3.1353\cdot 10^{-3}$ \\
   $w_6$ & $7.6882\cdot 10^{-4}$ & $6.9243\cdot 10^{-4}$ & $7.1265\cdot 10^{-4}$ & $6.8436\cdot 10^{-4}$ & $6.9083\cdot 10^{-4}$ & $6.8738\cdot 10^{-4}$ & $6.8749\cdot 10^{-4}$ & $6.8750\cdot 10^{-4}$ \\
   $w_7$ & $2.9747\cdot 10^{-1}$ & $2.9711\cdot 10^{-1}$ & $2.9656\cdot 10^{-1}$ & $2.9613\cdot 10^{-1}$ & $2.9591\cdot 10^{-1}$ & $2.9586\cdot 10^{-1}$ & $2.9585\cdot 10^{-1}$ & $2.9585\cdot 10^{-1}$ \\
   $w_8$ & $7.6882\cdot 10^{-4}$ & $6.9243\cdot 10^{-4}$ & $7.1265\cdot 10^{-4}$ & $6.8436\cdot 10^{-4}$ & $6.9083\cdot 10^{-4}$ & $6.8738\cdot 10^{-4}$ & $6.8749\cdot 10^{-4}$ & $6.8750\cdot 10^{-4}$ \\
   $w_9$ & $2.9747\cdot 10^{-1}$ & $2.9711\cdot 10^{-1}$ & $2.9656\cdot 10^{-1}$ & $2.9613\cdot 10^{-1}$ & $2.9591\cdot 10^{-1}$ & $2.9586\cdot 10^{-1}$ & $2.9585\cdot 10^{-1}$ & $2.9585\cdot 10^{-1}$ \\
$w_{10}$ & $7.7875\cdot 10^{-2}$ & $7.8013\cdot 10^{-2}$ & $7.8043\cdot 10^{-2}$ & $7.8170\cdot 10^{-2}$ & $7.8219\cdot 10^{-2}$ & $7.8352\cdot 10^{-2}$ & $7.8355\cdot 10^{-2}$ & $7.8357\cdot 10^{-2}$ \\
\hline
\end{tabular}
\caption{\label{tabconv}
The convergence of the electronic energy $\langle \Psi | \hat H | \Psi \rangle$ and the shape-block contributions $\{ w_i \}$ to $\Psi$ with the length $\mathcal{N}$ of the expansion~(\ref{a1}).
}
}
\end{table}
\addtolength{\tabcolsep}{4pt}

The availability of highly accurate approximations for the electronic wavefunction of the $^4 P_u$ quartet state of the lithium atom paves the way for the analysis of its bosonic content, as prescribed by the expansion (\ref{decomp}) and the inversion formula~(\ref{schemeinv}).  This analysis begins with the observation that the aggregate block functions
\begin{equation}
\Xi_k :=  \Xi_k(v_0) :=
 \frac{a_k}{36} \, \sum_{j=0}^{35} \bar\eta_k(\sigma_j) \Psi(\sigma_j v_0),
\quad  k = 0,...,10,
\label{a3}
\end{equation} 
where
\begin{equation}
a_k = \left[ \frac{1}{36} \sum_{j=0}^{35} \bar\eta_k(\sigma_j) \langle \Psi(v_0) |  \Psi(\sigma_j v_0) \rangle \right]^{-1/2}, \quad k = 0,...,10,
\label{a4}
\end{equation} 
are orthonormal, i.e.,
\begin{equation}
\begin{split}
\langle\Xi_k| \Xi_{k'} \rangle & = \frac{a_k a_{k'}}{36^2}  \sum_{j=0}^{35}  \sum_{j'=0}^{35}   
\bar\eta_k(\sigma_j)  \bar\eta_{k'}(\sigma_{j'})  \langle \Psi(\sigma_j v_0) |  \Psi(\sigma_{j'} v_0) \rangle
\\
& = \frac{a_k a_{k'}}{36^2}  \sum_{j=0}^{35}  \sum_{j'=0}^{35}   \bar\eta_k(\sigma_j)  \bar\eta_{k'}(\sigma_{j'}) 
\langle \Psi(v_0) |  \Psi(\sigma_j^{-1} \sigma_{j'} v_0) \rangle
\\
& = \frac{a_k a_{k'}}{36^2}  \sum_{J=0}^{35}  \langle \Psi(v_0) |  \Psi(\sigma_J v_0) \rangle  
\sum_{j=0}^{35}   \bar\eta_k(\sigma_j)  \bar\eta_{k'}(\sigma_{j}\sigma_{J})  \rangle
\\
& =  \delta_{kk'}  \frac{a_k a_{k'}}{36}  \sum_{J=0}^{35}  \bar\eta_k(\sigma_J)  \langle \Psi(v_0) |  \Psi(\sigma_J v_0) \rangle
= \delta_{kk'} , \quad 0 \le k,k' \le 10 ,
\end{split}
\label{a5}
\end{equation} 
per the character multiplication rule
\begin{equation}
 \frac{1}{36} \sum_{j=0}^{35}   \bar\eta_k(\sigma_j)  \bar\eta_{k'}(\sigma_{j}\sigma_{J})  = \delta_{kk'}  \bar\eta_k(\sigma_J), \quad 0 \le k,k' \le 10,
\label{a6}
\end{equation} 
that can be verified directly with the data in Table~\ref{deltamat}. Moreover, it follows from Eqs.~(\ref{schemeinv}) and (\ref{a3}) that they form a complete basis set for $\Psi$, i.e.
\begin{equation}
\Psi = \sum_{k=0}^{10} a_k \, \Xi_k .
\label{a7}
\end{equation} 
Consequently, the square $w_k = a_k^2$ of the real-valued amplitude $a_k$ quantifies the contribution of the $k$th shape block to $\Psi$.  Thanks to the employment of the ECGs in the expansion~(\ref{a1}), computation of these quantities is straightforward.

The bosonic densities
\begin{equation}
\mathcal{D}_i(\mathbf{r}_1) = \int \hspace{-4pt} \int |\Phi_i(\mathbf{r}_1,\mathbf{r}_2,\mathbf{r}_3)|^2  d\mathbf{r}_2 d\mathbf{r}_3, \quad i = 0,...,35,
\label{a8}
\end{equation} 
and the one-electron density,
\begin{equation}
\rho(\mathbf{r}_1) = 3\int \hspace{-4pt} \int |\Psi(\mathbf{r}_1,\mathbf{r}_2,\mathbf{r}_3)|^2  d\mathbf{r}_2 d\mathbf{r}_3,
\label{a9}
\end{equation} 
are also of interest. In general, their computation involves numerical quadratures. E.g., in the case of 
$\mathcal{D}_{32}(\mathbf{r}_1)$ that pertains to the shape $S_{32} = (x_1-x_2)(x_1-x_3)(x_2-x_3)$, integrals of the type
\[
\begin{split}
\int_{-\infty}^{\infty} \int_{-\infty}^{\infty} &\frac{F(x_1,x_2,x_3)}{9  (x_1-x_2)^2  (x_1-x_3)^2  (x_2-x_3)^2}  dx_2   dx_3 
\\& = \int _0^{\infty} \hspace{-1pt} \left[ \int_0^{2\pi} \frac{F(x_1,x_1+r \sin \phi, x_1+r \cos \phi)}{9  \sin^2 \phi \,\cos^2 \phi \,  (\sin \phi - \cos \phi)^2}  d\phi \right]   r^{-5}   dr,
\end{split}
\]
where $F(x_1,x_2,x_3)$ is a combination of  exponential functions with arguments given by quadratic forms in $x_1$, $x_2$, and $x_3$, have to be evaluated.

\begin{figure}
\begin{tabular}{ll}
(a) & (b) \\
\includegraphics[height=7cm]{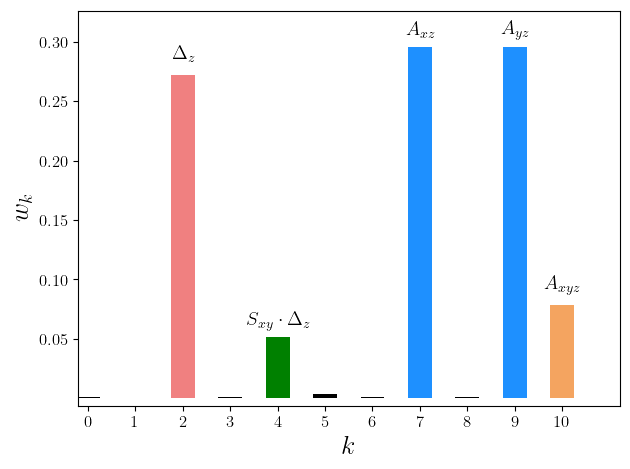}
& \includegraphics[height=7cm]{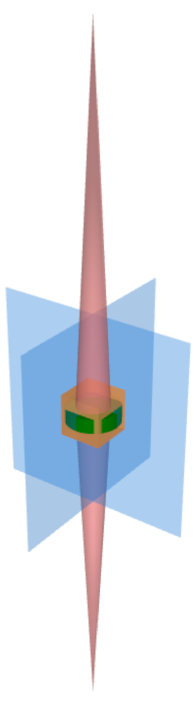}
\end{tabular}
\caption{\label{fig_blocks} (a) The shape-block contributions $w_k$ to the electronic wave function of the $^4P_u$ quartet state of the lithium atom. The labels above the bars indicate the type of shapes within the block (e.g., $S_{xy}$ means symmetric with respect to particle interchange and confined to the $xy$ plane; $A$ means antisymmetric, and $\Delta_z$ is the Vandermonde form in the $z$-direction). (b) Three-dimensional histogram of the contributions in (a) with the same color coding. The surfaces of the geometric figures are proportional to the probabilities. The yellow cube and green cylinder, the blue squares, and the red cones depict the three-, two-, and one-dimensional shape blocks, respectively.}
\end{figure}

\begin{figure}
\begin{center}
\begin{tabular}{lcl}
(a) &&  (b) \\
\includegraphics[width=0.4\linewidth]{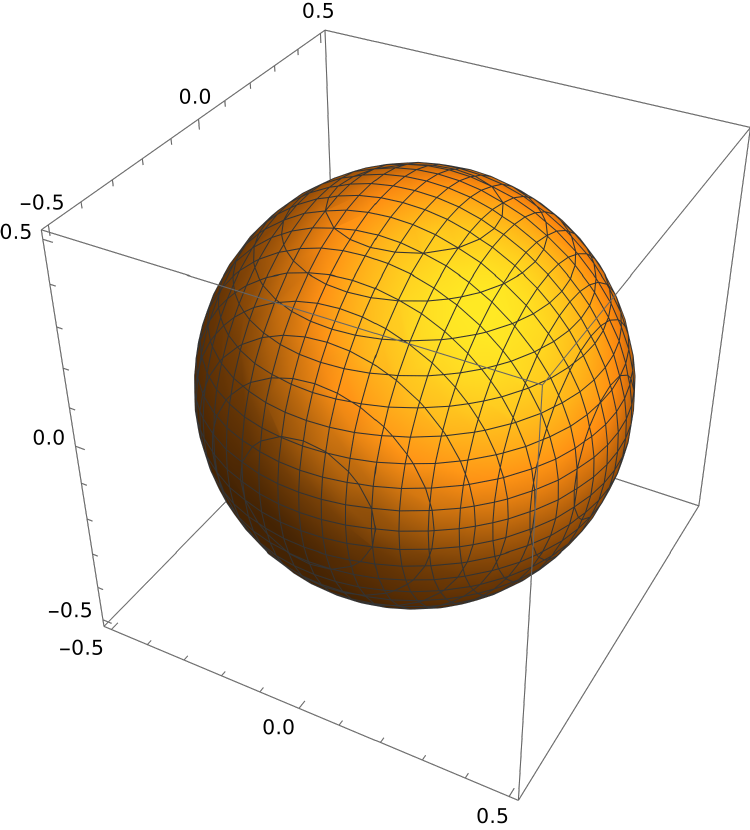}
&& \includegraphics[width=0.4\linewidth]{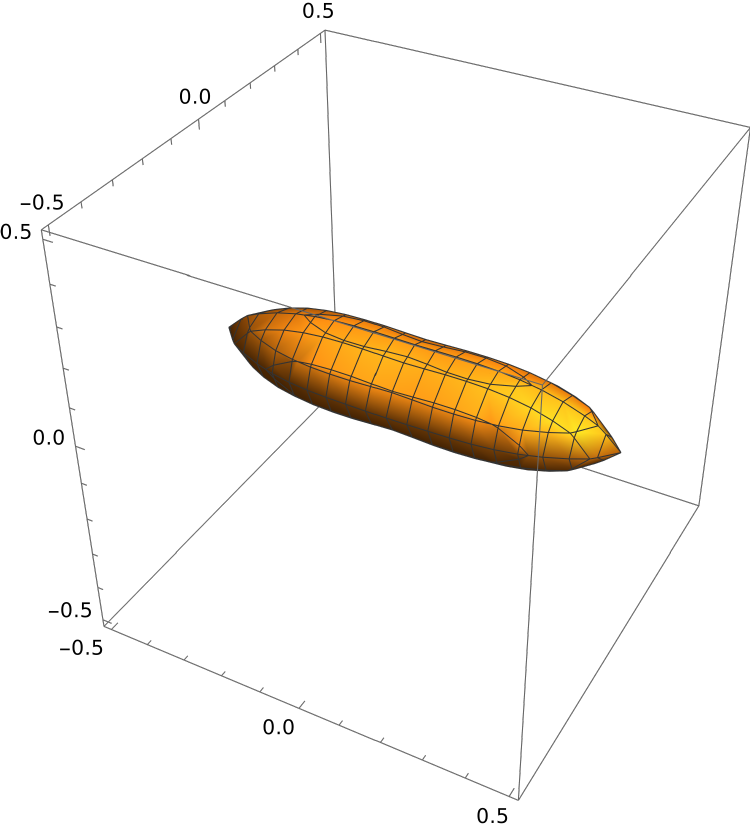}
\\ \multicolumn{1}{c}{$\rho(\mathbf{r}_1)$}
&&  \multicolumn{1}{c}{$\mathcal{D}_{32}(\mathbf{r}_1)$}
\\ (c) && (d) 
\\ \includegraphics[width=0.4\linewidth]{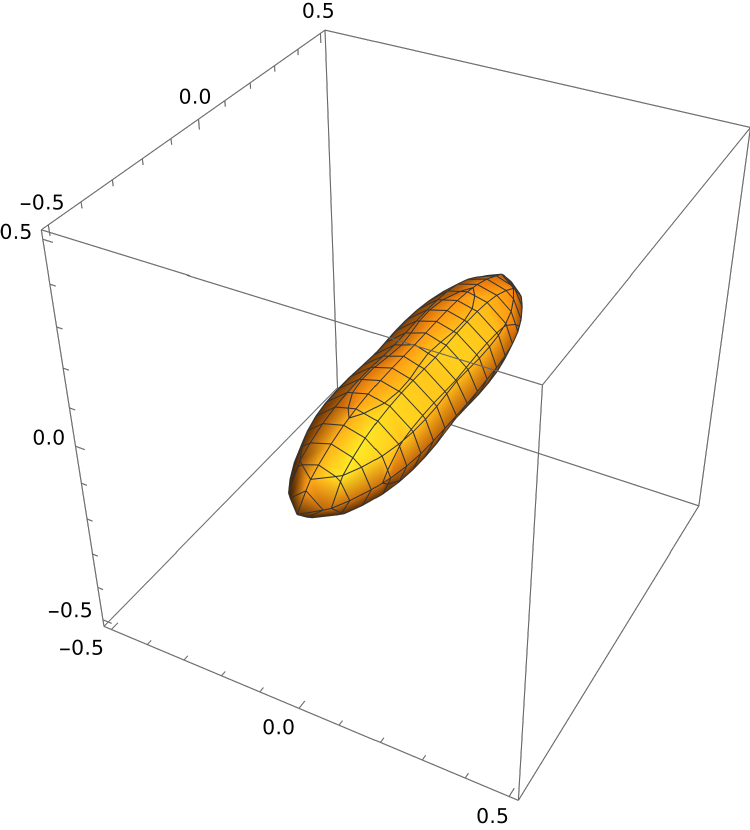}
&& \includegraphics[width=0.4\linewidth]{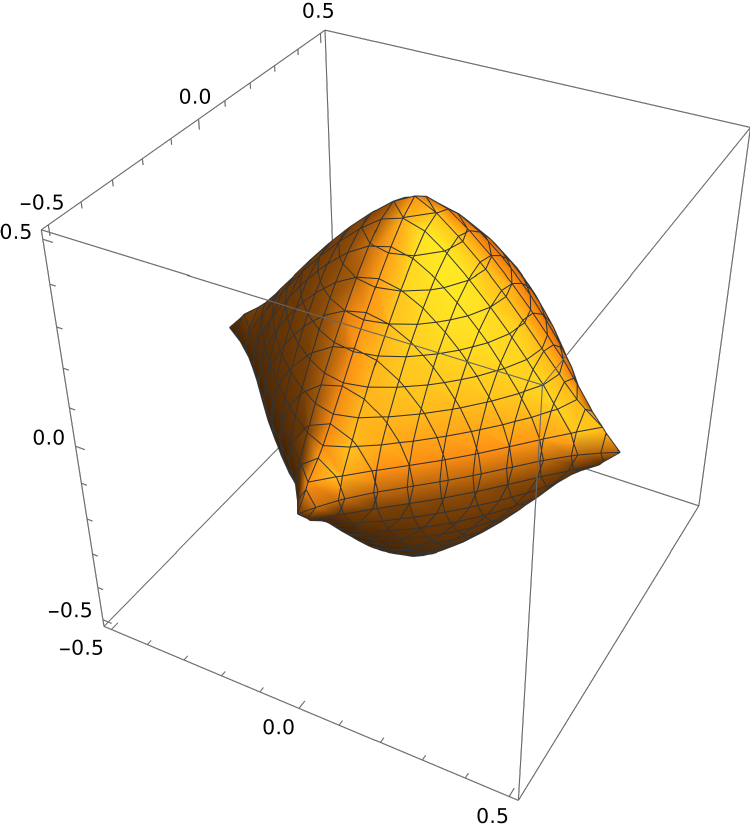}
\\
\multicolumn{1}{c}{$\mathcal{D}_{26}(\mathbf{r}_1)$} && \multicolumn{1}{c}{$\mathcal{D}_{23}(\mathbf{r}_1)$}
\end{tabular}
\end{center}
\caption{\label{fig_bosons} Isodensity contours for (a) the one-electron density $\rho(\mathbf{r}_1)$ and the bosonic densities (b) $\mathcal{D}_{32}(\mathbf{r}_1)$, (c) $\mathcal{D}_{26}(\mathbf{r}_1)$, and (d) $\mathcal{D}_{23}(\mathbf{r}_1)$ pertaining to the electronic wave function of the $^4P_u$ quartet state of the lithium atom. The contours that fit the cubic cages with $-\frac{1}{2}\le x_1,y_1,z_1\le \frac{1}{2}$ displayed in (a--d) correspond to the density values of $5.0\times 10^{-1}$, $9.0\times 10^{-5}$, $9.0\times 10^{-5}$, and $4.0\times 10^{-3}$, respectively.} 
\end{figure}

Inspection of Table~\ref{tabconv} reveals the rapid convergence of the shape-block contributions $\{ w_i \}$ to $\Psi$ with the length $\mathcal{N}$ of the expansion~(\ref{a1}) that rivals that of the electronic energy $\langle \Psi | \hat H | \Psi \rangle$.  Although surprising at first glance, it demonstrates the relative insensitivity of these contributions to the behavior of $\Psi$ in the vicinities of the electron-electron coalescence points that is poorly reproduced by the ECGs.  However, the most important feature of the data compiled in Table~\ref{tabconv} is the predominance of just three shape blocks, namely the 2nd, 7th, and 9th, that together contribute over 86\% to $\Psi$.  Their next two most important counterparts (namely the 4th and 10th ones) contribute almost 13\%, leaving only ca.\ 0.6\% for the remaining 6 shape blocks. To be explicit, $99.4$\% of the spectral weight of the practically exact Li wave function $\Psi$ expressed in our basis of $1278$ correlated Gaussians can be meaningfully rewritten as
\begin{equation}
\Psi\approx 0.27208\,\Xi_2 + 0.051493\,\Xi_4
+ 0.29585\,\Xi_7 +  0.29585\,\Xi_9 + 0.078357\,\Xi_{10},
\end{equation}
where the block functions $\Xi_k$~\eqref{a3} are fixed (Fourier-like) transforms of $\Psi$. In other words, the wave function with extremely complex dependencies on the  positions of electrons can be condensed with only a minor loss of information to 5 blocks comprising a total of 21 shapes---a truly remarkable compact description (Fig.~\ref{fig_blocks}). As expected, the contributions conform to the symmetry of $\Psi$ (i.e., $w_0=w_3$, $w_6=w_8$, and $w_7=w_9$).   
Due to the presence of the core $1s$ orbital in the Slater determinant that provides the zeroth-order approximation of $\Psi$, the one-electron density $\rho(\mathbf{r}_1)$ is almost spherically symmetric at short distances from the nucleus (Fig.~\ref{fig_bosons}a). In contrast, the bosonic densities exhibit complicated angular dependences. For example, the equidensity contours of the symmetry-equivalent 
$\mathcal{D}_{32}(\mathbf{r}_1)$ and $\mathcal{D}_{26}(\mathbf{r}_1)$  are cigar-shaped (Figs.~\ref{fig_bosons}b and~\ref{fig_bosons}c), whereas those of  
$\mathcal{D}_{23}(\mathbf{r}_1)$ have shapes resembling two joined tetragonal-pyramid variants of \emph{zongzi} rice cakes (Fig.~\ref{fig_bosons}d). As expected from the discussion in Appendix~\ref{altclass}, $\mathcal{D}_{23}(\mathbf{r}_1)$ does not have cylindrical symmetry.

The overall computational complexity of the scheme is dominated by the number $D:=(N!)^{d-1}$ of shapes of $N$ fermions in $d$ dimensions. In principle, they need to be generated once and for all for each combination of $N$ and $d$. The construction of the transformation matrix $F$ requires $D$ evaluations per character. The number of distinct characters of $\mathcal{S}_N^{d-1}$ is $p(N)^{d-1}\sim\exp(\sim (d-1)\sqrt{N})<<D$ for large $N$, where $p(N)$ is the number of partitions of $N$. Because of this mismatch in magnitudes, the size of the subspaces extracted by the characters is expected to increase. The key unknown is whether the number of shape blocks will behave in a similar fashion, or remain manageable with the shape blocks themselves increasing in size. Given the qualitative classification of shape blocks obtained here, the latter is expected, indeed, their number could remain equal to eleven. Finally, in order to obtain the bosonic coefficient functions, one needs to compute the wave function with the same number $D$ of permutations of its arguments. This cost scales as $D\cdot B$, with $B$ being the cost of evaluating the wave function. If we collect all factors multiplying $D$ into some $C$, the latter is logarithmically negligible, $\log C\ll\log D$, for larger values of $N$. For cases where $D$ is prohibitively large, an alternative stochastic approach has been envisaged~\cite{Sunko20}.

\section{Discussion and conclusion}\label{sec12}

The description of finite systems by quantum theory has always been subject to some tension between the microscopic approach, which commences with constituent fermions, and the phenomenological approach that focuses on their collective properties. This tension was first overt in nuclear physics, where the liquid-drop (collective) and shell (fermionic) models coexisted for decades without one ever being fully reduced to the other~\cite{Bohr76,Mottelson76}. When the number of fermions is very small, one may well ask whether the term ``collective'' retains a physical meaning at all.

Specifically, imagine a collective excited state of a finite system as a superposition of particle-hole excitations over a Fermi sea,
\begin{equation}
a^\dagger_1\cdots a^\dagger_na_{n+1}\cdots a_{2n}\left|F\right>.
\label{standard}
\end{equation}
One would like to gather all such terms into an effective boson $b^\dagger$, so that $(b^\dagger)^M\left|F\right>$ describes the $M$-th excited state of a collective rotation or vibration band. Despite the considerable phenomenological success of some such approaches~\cite{Janssen74}, all based on Lie-group theory, this program remained dependent on choices of subgroup decompositions and fermion coupling schemes that could not be justified microscopically~\cite{Iachello19}.

These considerations have been brought to a head by the advent of manufactured finite systems, such as few-electron quantum dots~\cite{Bakker15} or Rydberg atoms~\cite{Wu21}, that hold the promise of becoming the physical building blocks of quantum computers. Robustness, already mentioned in the Introduction, can be described in the collective-model language as a distinction between excitation bands and their bandheads, i.e., the lowest-lying states of each band. In deexcitation cascades, states easily populated by decay of higher states are in the same band, while transitions between bands are much weaker and often suppressed by selection rules (i.e., due to their different symmetry). The long-standing intuitive understanding of this behavior is that excitations within each band are bosonic in character, while the bands differ in the spatial configuration of the fermionic wave function~\cite{Bohr76,Mottelson76,Janssen74,Iachello19}.

Here, we have applied a precise algebraic definition of the collective excitation to the lowest quartet state of lithium, with three electrons in the highest-spin configuration. The key breakthrough is the formal separation of kinematics from  dynamics, based on the theory of invariants of the kinematic discrete alternating group $\mathcal{A}_N$~\eqref{alter}, rather than of continuous (Lie) groups related to dynamical symmetries~\cite{Iachello19}. By listing explicitly the limited number of ways in which a many-fermion wave function can satisfy the Pauli principle, namely the shapes, dynamical bosonic excitations are defined unambiguously as their symmetric-function coefficients. The new feature of the scheme~\eqref{scheme} is that it is additive over the shapes, each of which can be regarded as a potential ground state, or bandhead, rather than multiplicative with an assumed ground state as in~\eqref{standard}. Our transform of the $N$-fermion wave function by characters of the symmetric group $\mathcal{S}_N^2$ is the natural finite-system analogue of the Fourier transform for infinite systems. Thus, we rigorously deduce collective modes directly from a microscopic fermionic wave function.

The one-electron densities of the bosonic functions displayed in Fig.~\ref{fig_bosons} are the first-ever depiction of such collective modes. They readily defy some simplistic expectations. Thus, the density of the bosonic function $\Phi_{23}$ is nearly permutationally symmetric among the axes despite the antisymmetry of the corresponding component $\Phi_{23}\Delta_z$ of the wave function being ensured by the highly anisotropic Vandermonde factor $\Delta_z$. Most remarkably, the non-trivial spatial distributions of the bosonic functions are hidden behind a nearly featureless electron density, showing the loss of readily extractable information in the latter.

This analysis is called nonlinear~\cite{Breiding23,Michalek21}, in contrast to the linear analysis of the $c$-number amplitudes in a standard antisymmetrized basis. Its advantage stems from the reduction of hundreds, potentially thousands, of basis vectors to just a handful of meaningful numbers encoding the configuration of the wave function (as in Fig.~\ref{fig_blocks}). Thus, this analysis provides a well-defined way to specify requirements for bespoke configurations to be used in quantum computations. For example, to construct a state most different from the one in Fig.~\ref{fig_blocks} (as, e.g., orthohydrogen is different from parahydrogen), one would require blocks $3$, $5$, and $8$ to be dominant. Conversely, and more broadly, the nonlinear approach is a prescription for writing a variational wave function which encompasses the whole Hilbert space in a finite number of terms.

The ortho- vs.\ parahydrogen paradigm brings us to questions of superselection and entanglement~\cite{Ding23,Ding24,Aliverti-Piuri24}. The original superselection rule means physically that a pure quantum state cannot be prepared in a coherent superposition of integer and half-integer angular momenta~\cite{Wick52}. That understanding of superselection appears too narrow today. Applying the principle of superselection to general superpositions leads to a significant reassessment of their quantum-information content~\cite{Ding21}.

Shapes are a manifestation of the intrinsic structure of many-body wave functions, which introduces an absolute measure of their complexity, based on their derivative structure, with a computable definition implemented in Ref.~\cite{Sunko22} for the $36$ shapes studied here. Different molecules with the same chemical formula, such as chemical isomers, behave like distinguishable systems, so they are superselected in a practical sense of the word by zero tunnelling amplitudes, even in the absence of an incompatible underlying symmetry. Conversely, even much larger molecules of the same kind, like fullerenes, can exhibit interference patterns~\cite{Brezger02}. Ortho- and parahydrogen occupy an interesting middle ground in this sense, because they differ in the nuclear wave function of the two protons, while having the same arrangement of atoms in space. Shapes provide a natural framework for understanding the emergence of superselection in configuration space.

We regard this application of the superselection paradigm as practical because it does not rely on exactly conserved algebraic operators, but on an adiabatic topological invariant that can be violated by non-adiabatic processes (e.g., tunneling) that are presumed to be rare. Specifically, the total charge is an exact superselection operator in the usual sense. On the other hand, the charge enclosed by a nodal surface of a many-body wave function is an adiabatic topological invariant, because an infinitely slowly moving surface on which the wave function vanishes cannot pass over an electron~\cite{Herzfeld49}. The only way that the enclosed charge can change is by nodal surfaces breaking up or merging, which are non-adiabatic processes, corresponding, e.g., to reconfiguration of isomers. It is clear from the analysis above that different shape blocks have different nodal structures, hence they define superselection sectors in this practical sense. It is also clear that non-adiabatic reconfiguration is less probable with more complicated shapes, i.e., when the number of particles increases.

The somewhat surprising incompatibility of shapes with classification by angular momentum, described in Appendix~\ref{altclass}, provides further evidence of their fundamental nature. It is a trivial observation that one must excite a state in order to study it. It is less trivial that the excitation process itself imposes the rotation symmetry of laboratory space onto the atom, while the underlying Pauli kinematics of many-body Hilbert space conforms to the more primitive permutation symmetry of directions in space. This point of view sheds an interesting light on the well-known restoration of orbital angular momentum of open-$d$-shell ions at points of cubic symmetry in a crystal lattice~\cite{Khan19}. Apparently, degeneracy due to the permutational symmetry of the crystal axes is indistinguishable from that due to rotation symmetry.

A generalization to less-than-highest spin states is easily envisaged. Components of the wave function appear factorized by projection,
\begin{equation}
\left[\hat{\mathcal{A}}\Psi(1)\cdots \Psi(K)\right]_\uparrow
\left[\hat{\mathcal{A}}\Psi(K+1)\cdots \Psi(N)\right]_\downarrow,
\end{equation}
where the antisymmetrized space factors can be expressed by shapes, separately for each projection. Because the number of shapes scales with $N!^2$, the zero-projection ($K=N/2$) subspace requires significantly less work than the highest-spin subspace. Even in the worst-case scenario, where the up- and down-spin efforts multiply,
\begin{equation}
(N/2)!^4<N!^2 \Longleftrightarrow
1 < 4^N.
\end{equation}
If one can be reused (best case), $N^N$ appears in place of $4^N$.

Much remains to be done along the lines traced in this work. The systematization and automatization of the classification obtained here should enable its extension to at least $N=5$ fermions, corresponding to a half-open $d$ shell, thus accounting for transition-metal compounds that comprise a sizable part of functional materials today. Building on that, one should develop a multicenter formalism, which would enable the treatment of molecules and crystal lattices on equal footing with atoms.

In conclusion, we have presented a nonlinear structural analysis of the lowest-energy quartet state of the lithium atom in terms of fundamental many-body invariants of the Pauli principle. We hope that it will influence research on manufacturing electronic states with bespoke quantum properties.

\section*{Acknowledgements}

DKS gratefully acknowledges conversations with D.~Svrtan. The research described in this publication was funded by the Croatian Science Foundation under Project IP-2022-10-3382 and by the National Science Center (Poland) under grant 2022/47/B/ST4/00002.

\bibliographystyle{unsrturl}
\bibliography{lithium_accepted.bib}

\onecolumn
\appendix

\section{Conventions and tables}\label{conventab}

The group $\mathcal{S}_3$ consists of six permutations, in cycle notation:
\begin{equation}
\mathcal{S}_3 = [(1)(2)(3), (23)(1), (12)(3), (13)(2), (123), (132)].
\end{equation} 
When understood as permuting three objects $(x,y,z)$, it acts irreducibly in the two-dimensional subspace (E) orthogonal to the symmetric (S) invariant $x+y+z$:
\begin{equation}
\begin{pmatrix}
x + y + z\\
x-y\\
x+y-2z
\end{pmatrix}
=\begin{pmatrix*}[r]
1 & 1 & 1 \\
1 & -1 & 0 \\
1 & 1 & -2
\end{pmatrix*}\begin{pmatrix}
x\\
y\\
z
\end{pmatrix}.
\end{equation} 
We require that the first vector spanning the E-subspace be antisymmetric and the second symmetric with respect to the transposition $x\leftrightarrow y$, and that the transformation matrix be orthogonal by rows. In the so-defined subspace, the group $\mathcal{S}_3$ is represented by the list of matrices:
\begin{equation}
\begin{split}
E =&\left[
\begin{pmatrix*}[r]
1 & 0\\
0 & 1
\end{pmatrix*},\begin{pmatrix*}[r]
1/2 & 1/2\\
3/2 & -1/2
\end{pmatrix*},\begin{pmatrix*}[r]
-1 & 0\\
0 & 1
\end{pmatrix*},
\right. \\ &\left.
\begin{pmatrix*}[r]
1/2 & -1/2\\
-3/2 & -1/2
\end{pmatrix*},
\begin{pmatrix*}[r]
-1/2 & 1/2\\
-3/2 & -1/2
\end{pmatrix*},
\begin{pmatrix*}[r]
-1/2 & -1/2\\
3/2 & -1/2
\end{pmatrix*}
\right].
\end{split}
\end{equation} 
They act on columns from the left, e.g., to transpose $y$ and $z$:
\begin{equation}
\begin{pmatrix*}[r]
1/2 & 1/2\\
3/2 & -1/2
\end{pmatrix*}
\begin{pmatrix}
x-y \\
x+y-2z
\end{pmatrix}
=\begin{pmatrix}
x-z \\
x+z-2y
\end{pmatrix}.
\end{equation} 
Another list of matrices gives rise to the same group multiplication table:
\begin{equation}
\begin{split}
\bar{E} =&\left[
\begin{pmatrix*}[r]
1 & 0\\
0 & 1
\end{pmatrix*},\begin{pmatrix*}[r]
1/2 & 3/2\\
1/2 & -1/2
\end{pmatrix*},\begin{pmatrix*}[r]
-1 & 0\\
0 & 1
\end{pmatrix*},
\right. \\ &\left.
\begin{pmatrix*}[r]
1/2 & -3/2\\
-1/2 & -1/2
\end{pmatrix*},
\begin{pmatrix*}[r]
-1/2 & 3/2\\
-1/2 & -1/2
\end{pmatrix*},
\begin{pmatrix*}[r]
-1/2 & -3/2\\
1/2 & -1/2
\end{pmatrix*}
\right].
\end{split}
\end{equation} 
They are related to the former by exchanging the $1$ and $3$ on the skew diagonals. In order to obtain the same group action, act on rows from the right \emph{and} read the cycles backwards, which is the same as $(123)$ and $(132)$ exchanging places. The difference between $E$ and $\bar{E}$ is invisible to character theory, because the permutation of group elements induced by reading cycles backwards occurs within conjugacy classes. Both matrix representations appear naturally when classifying the shapes, so we distinguish them in Table~\ref{evalQ} below. In particular, $\bar{E}$ appears together with antisymmetric ($A$) singlets, analogously to axial vectors in angular momentum theory.

In the following tables, we list shapes, variable sets, and characters of $\mathcal{S}_3^2$ in the order in which they appear in the text.

\begin{table}[H]
\begin{center}
\begin{tabular}{|r|r|}
\hline
$v_{0} = (\ldots, y_1, y_2, y_3, z_1, z_2, z_3)$ &
$v_{1} = (\ldots, y_1, y_2, y_3, z_1, z_3, z_2)$  \\
$v_{2} = (\ldots, y_1, y_2, y_3, z_2, z_1, z_3)$ &
$v_{3} = (\ldots, y_1, y_2, y_3, z_3, z_2, z_1)$  \\
$v_{4} = (\ldots, y_1, y_2, y_3, z_2, z_3, z_1)$ &
$v_{5} = (\ldots, y_1, y_2, y_3, z_3, z_1, z_2)$  \\
$v_{6} = (\ldots, y_1, y_3, y_2, z_1, z_2, z_3)$ &
$v_{7} = (\ldots, y_2, y_1, y_3, z_1, z_2, z_3)$  \\
$v_{8} = (\ldots, y_3, y_2, y_1, z_1, z_2, z_3)$ &
$v_{9} = (\ldots, y_1, y_3, y_2, z_1, z_3, z_2)$  \\
$v_{10} = (\ldots, y_2, y_1, y_3, z_1, z_3, z_2)$ &
$v_{11} = (\ldots, y_3, y_2, y_1, z_1, z_3, z_2)$  \\
$v_{12} = (\ldots, y_1, y_3, y_2, z_2, z_1, z_3)$ &
$v_{13} = (\ldots, y_1, y_3, y_2, z_3, z_2, z_1)$  \\
$v_{14} = (\ldots, y_2, y_1, y_3, z_2, z_1, z_3)$ &
$v_{15} = (\ldots, y_2, y_1, y_3, z_3, z_2, z_1)$  \\
$v_{16} = (\ldots, y_3, y_2, y_1, z_2, z_1, z_3)$ &
$v_{17} = (\ldots, y_3, y_2, y_1, z_3, z_2, z_1)$  \\
$v_{18} = (\ldots, y_1, y_3, y_2, z_2, z_3, z_1)$ &
$v_{19} = (\ldots, y_2, y_1, y_3, z_2, z_3, z_1)$  \\
$v_{20} = (\ldots, y_3, y_2, y_1, z_2, z_3, z_1)$ &
$v_{21} = (\ldots, y_1, y_3, y_2, z_3, z_1, z_2)$  \\
$v_{22} = (\ldots, y_2, y_1, y_3, z_3, z_1, z_2)$ &
$v_{23} = (\ldots, y_3, y_2, y_1, z_3, z_1, z_2)$  \\
$v_{24} = (\ldots, y_2, y_3, y_1, z_1, z_2, z_3)$ &
$v_{25} = (\ldots, y_3, y_1, y_2, z_1, z_2, z_3)$  \\
$v_{26} = (\ldots, y_2, y_3, y_1, z_1, z_3, z_2)$ &
$v_{27} = (\ldots, y_3, y_1, y_2, z_1, z_3, z_2)$  \\
$v_{28} = (\ldots, y_2, y_3, y_1, z_2, z_1, z_3)$ &
$v_{29} = (\ldots, y_2, y_3, y_1, z_3, z_2, z_1)$  \\
$v_{30} = (\ldots, y_3, y_1, y_2, z_2, z_1, z_3)$ &
$v_{31} = (\ldots, y_3, y_1, y_2, z_3, z_2, z_1)$  \\
$v_{32} = (\ldots, y_2, y_3, y_1, z_2, z_3, z_1)$ &
$v_{33} = (\ldots, y_3, y_1, y_2, z_2, z_3, z_1)$  \\
$v_{34} = (\ldots, y_2, y_3, y_1, z_3, z_1, z_2)$ &
$v_{35} = (\ldots, y_3, y_1, y_2, z_3, z_1, z_2)$ \\ \hline
\end{tabular}
\end{center}
\caption{The group $\mathcal{S}_3^2=\mathcal{S}_3\times \mathcal{S}_3$ comprises $36$ permutations $\sigma_0, \ldots, \sigma_{35}$, which act separately on the $y$ and $z$ coordinates in the list $v_0=(x_1, x_2, x_3, y_1, y_2, y_3, z_1, z_2, z_3)$. We identify them by their action, i.e., $v_i = \sigma_iv_0$. The ellipsis $\ldots$ stands for $x_1, x_2, x_3$.
\label{varsSS3}}
\end{table}

\begin{table}[H]
\begin{center}
\begin{tabular}{|c|c|l|}
\hline
$S_{0}$ & $1$ & $x_{222}y_{222}z_{222}$\\ \hline
$S_{1}$ & $4$ & $2x_{212}y_{212}z_{222} + x_{221}y_{212}z_{222} + x_{212}y_{221}z_{222}+2x_{221}y_{221}z_{222}$\\
$S_{2}$ & $6$ & $2x_{212}y_{222}z_{212} + x_{221}y_{222}z_{212} + x_{212}y_{222}z_{221}+2x_{221}y_{222}z_{221}$\\
$S_{3}$ & $8$ & $2x_{222}y_{212}z_{212} + x_{222}y_{221}z_{212} + x_{222}y_{212}z_{221}+2x_{222}y_{221}z_{221}$\\ \hline
$S_{4}$ & $4$ & $-x_{121}y_{212}z_{222} + x_{211}y_{212}z_{222} + x_{121}y_{221}z_{222}+2x_{211}y_{221}z_{222}$\\
$S_{5}$ & $6$ & $-x_{121}y_{222}z_{212} + x_{211}y_{222}z_{212} + x_{121}y_{222}z_{221}+2x_{211}y_{222}z_{221}$\\
$S_{6}$ & $4$ & $-x_{212}y_{121}z_{222} + x_{221}y_{121}z_{222} + x_{212}y_{211}z_{222}+2x_{221}y_{211}z_{222}$\\
$S_{7}$ & $10$ & $-x_{221}y_{212}z_{212} - x_{212}y_{221}z_{212} - x_{221}y_{221}z_{212}$\\
&& $ \phantom{x_{121}y_{121}z_{121} + x_{211}} - x_{212}y_{212}z_{221} - x_{221}y_{212}z_{221} - x_{212}y_{221}z_{221}$\\
$S_{8}$ & $6$ & $-x_{212}y_{222}z_{121} + x_{221}y_{222}z_{121} + x_{212}y_{222}z_{211}+2x_{221}y_{222}z_{211}$\\
$S_{9}$ & $8$ & $-x_{222}y_{121}z_{212} + x_{222}y_{211}z_{212} + x_{222}y_{121}z_{221}+2x_{222}y_{211}z_{221}$\\
$S_{10}$ & $8$ & $-x_{222}y_{212}z_{121} + x_{222}y_{221}z_{121} + x_{222}y_{212}z_{211}+2x_{222}y_{221}z_{211}$\\ \hline
$S_{11}$ & $4$ & $2x_{121}y_{121}z_{222} + x_{211}y_{121}z_{222} + x_{121}y_{211}z_{222}+2x_{211}y_{211}z_{222}$\\
$S_{12}$ & $10$ & $-x_{121}y_{212}z_{212} - x_{211}y_{212}z_{212} - x_{211}y_{221}z_{212}$\\
&& $ \phantom{x_{121}y_{121}z_{121} + x_{211}} - x_{211}y_{212}z_{221} + x_{121}y_{221}z_{221}$\\
$S_{13}$ & $6$ & $2x_{121}y_{222}z_{121} + x_{211}y_{222}z_{121} + x_{121}y_{222}z_{211}+2x_{211}y_{222}z_{211}$\\
$S_{14}$ & $10$ & $-x_{212}y_{121}z_{212} - x_{212}y_{211}z_{212} - x_{221}y_{211}z_{212}$\\
&& $ \phantom{x_{121}y_{121}z_{121} + x_{211}} + x_{221}y_{121}z_{221} - x_{212}y_{211}z_{221}$\\
$S_{15}$ & $10$ & $-x_{212}y_{212}z_{121} + x_{221}y_{221}z_{121} - x_{212}y_{212}z_{211}$\\
&& $ \phantom{x_{121}y_{121}z_{121} + x_{211}} - x_{221}y_{212}z_{211} - x_{212}y_{221}z_{211}$\\
$S_{16}$ & $8$ & $2x_{222}y_{121}z_{121} + x_{222}y_{211}z_{121} + x_{222}y_{121}z_{211}+2x_{222}y_{211}z_{211}$\\ \hline
$S_{17}$ & $9$ & $3x_{210}y_{221}z_{212}-3x_{210}y_{212}z_{221}$\\
$S_{18}$ & $10$ & $-x_{121}y_{121}z_{212} + x_{211}y_{211}z_{212} - x_{121}y_{121}z_{221}$\\
&& $ \phantom{x_{121}y_{121}z_{121} + x_{211}} - x_{211}y_{121}z_{221} - x_{121}y_{211}z_{221}$\\
$S_{19}$ & $10$ & $-x_{121}y_{212}z_{121} - x_{121}y_{221}z_{121} - x_{211}y_{221}z_{121}$\\
&& $ \phantom{x_{121}y_{121}z_{121} + x_{211}} + x_{211}y_{212}z_{211} - x_{121}y_{221}z_{211}$\\
$S_{20}$ & $7$ & $3x_{221}y_{210}z_{212}-3x_{212}y_{210}z_{221}$\\
$S_{21}$ & $10$ & $-x_{212}y_{121}z_{121} - x_{221}y_{121}z_{121} - x_{221}y_{211}z_{121}$\\
&& $\phantom{x_{121}y_{121}z_{121} + x_{211}} - x_{221}y_{121}z_{211} + x_{212}y_{211}z_{211}$\\
$S_{22}$ & $5$ & $3x_{221}y_{212}z_{210}-3x_{212}y_{221}z_{210}$\\ \hline
$S_{23}$ & $2$ & $6x_{210}y_{210}z_{222}$ \\
$S_{24}$ & $9$ & $3x_{210}y_{121}z_{212} +3x_{210}y_{211}z_{212} +3x_{210}y_{121}z_{221}$ \\
$S_{25}$ & $9$ & $3x_{210}y_{212}z_{121} +3x_{210}y_{221}z_{121} +3x_{210}y_{212}z_{211}$ \\
$S_{26}$ & $3$ & $6x_{210}y_{222}z_{210}$ \\
$S_{27}$ & $7$ & $3x_{121}y_{210}z_{212} +3x_{211}y_{210}z_{212} +3x_{121}y_{210}z_{221}$ \\
$S_{28}$ & $10$ & $ -x_{211}y_{121}z_{121} - x_{121}y_{211}z_{121} - x_{211}y_{211}z_{121}$ \\
&& $\phantom{x_{121}y_{121}z_{121} + x_{211}} - x_{121}y_{121}z_{211} - x_{211}y_{121}z_{211} - x_{121}y_{211}z_{211}$ \\
$S_{29}$ & $5$ & $3x_{121}y_{212}z_{210} +3x_{211}y_{212}z_{210} +3x_{121}y_{221}z_{210}$ \\
$S_{30}$ & $7$ & $3x_{212}y_{210}z_{121} +3x_{221}y_{210}z_{121} +3x_{212}y_{210}z_{211}$ \\
$S_{31}$ & $5$ & $3x_{212}y_{121}z_{210} +3x_{221}y_{121}z_{210} +3x_{212}y_{211}z_{210}$ \\
$S_{32}$ & $0$ & $6x_{222}y_{210}z_{210}$ \\ \hline
$S_{33}$ & $9$ & $ -3x_{210}y_{211}z_{121} +3x_{210}y_{121}z_{211}$ \\
$S_{34}$ & $7$ & $ -3x_{211}y_{210}z_{121} +3x_{121}y_{210}z_{211}$ \\
$S_{35}$ & $5$ & $ -3x_{211}y_{121}z_{210} +3x_{121}y_{211}z_{210}$ \\ \hline
\end{tabular}
\end{center}
\caption{The shapes are listed in terms of the harmonic polynomials~\eqref{cvf}. Each term appears in exactly one shape. Block membership according to Table~\ref{blocktable} is indicated in the second column. Shapes of the same total degree are separated by lines.
\label{evalS}}
\end{table}

\addtolength{\tabcolsep}{-4pt}
\begin{sidewaystable}
\begin{tabular}{cc|rrrrrrrrrrrrrrrrrrrrrrrrrrrrrrrrrrrr}
$k$ & $j$: & \phantom{0}0 &  \phantom{0}1 &  \phantom{0}2 &  \phantom{0}3 &  \phantom{0}4 &  \phantom{0}5 &  \phantom{0}6 &  \phantom{0}7 &  \phantom{0}8 &  \phantom{0}9 & 10 & 11 & 12 & 13 & 14 & 15 & 16 & 17 & 18 & 19 & 20 & 21 & 22 & 23 & 24 & 25 & 26 & 27 & 28 & 39 & 30 & 31 & 32 & 33 & 34 & 35 \\ \hline
0 & & 1 & 1 & 1 & 1 & 1 & 1 & 1 & 1 & 1 & 1 & 1 & 1 & 1 & 1 & 1 & 1 & 1 & 1 & 1 & 1 & 1 & 1 & 1 & 1 & 1 & 1 & 1 & 1 & 1 & 1 & 1 & 1 & 1 & 1 & 1 & 1 \\
1 & & 1 & -1 & -1 & -1 & 1 & 1 & -1 & -1 & -1 & 1 & 1 & 1 & 1 & 1 & 1 & 1 & 1 & 1 & -1 & -1 & -1 & -1 & -1 & -1 & 1 & 1 & -1 & -1 & -1 & -1 & -1 & -1 & 1 & 1 & 1 & 1 \\
2 & & 1 & -1 & -1 & -1 & 1 & 1 & 1 & 1 & 1 & -1 & -1 & -1 & -1 & -1 & -1 & -1 & -1 & -1 & 1 & 1 & 1 & 1 & 1 & 1 & 1 & 1 & -1 & -1 & -1 & -1 & -1 & -1 & 1 & 1 & 1 & 1 \\
3 & & 1 & 1 & 1 & 1 & 1 & 1 & -1 & -1 & -1 & -1 & -1 & -1 & -1 & -1 & -1 & -1 & -1 & -1 & -1 & -1 & -1 & -1 & -1 & -1 & 1 & 1 & 1 & 1 & 1 & 1 & 1 & 1 & 1 & 1 & 1 & 1 \\
4 & & 2 & -2 & -2 & -2 & 2 & 2 & 0 & 0 & 0 & 0 & 0 & 0 & 0 & 0 & 0 & 0 & 0 & 0 & 0 & 0 & 0 & 0 & 0 & 0 & -1 & -1 & 1 & 1 & 1 & 1 & 1 & 1 & -1 & -1 & -1 & -1 \\
5 & & 2 & 2 & 2 & 2 & 2 & 2 & 0 & 0 & 0 & 0 & 0 & 0 & 0 & 0 & 0 & 0 & 0 & 0 & 0 & 0 & 0 & 0 & 0 & 0 & -1 & -1 & -1 & -1 & -1 & -1 & -1 & -1 & -1 & -1 & -1 & -1 \\
6 & & 2 & 0 & 0 & 0 & -1 & -1 & -2 & -2 & -2 & 0 & 0 & 0 & 0 & 0 & 0 & 0 & 0 & 0 & 1 & 1 & 1 & 1 & 1 & 1 & 2 & 2 & 0 & 0 & 0 & 0 & 0 & 0 & -1 & -1 & -1 & -1 \\
7 & & 2 & 0 & 0 & 0 & -1 & -1 & 2 & 2 & 2 & 0 & 0 & 0 & 0 & 0 & 0 & 0 & 0 & 0 & -1 & -1 & -1 & -1 & -1 & -1 & 2 & 2 & 0 & 0 & 0 & 0 & 0 & 0 & -1 & -1 & -1 & -1 \\
8 & & 4 & 0 & 0 & 0 & -2 & -2 & 0 & 0 & 0 & 0 & 0 & 0 & 0 & 0 & 0 & 0 & 0 & 0 & 0 & 0 & 0 & 0 & 0 & 0 & -2 & -2 & 0 & 0 & 0 & 0 & 0 & 0 & 1 & 1 & 1 & 1 \\
\end{tabular}
\vskip 1ex
\caption{The characters $\chi_k(\sigma_j)$ of the group $\mathcal{S}_3^2$, with $\sigma_j$ ordered as defined by Table~\ref{varsSS3}.
\label{varcharsSS3}}
\end{sidewaystable}
\addtolength{\tabcolsep}{4pt}

\addtolength{\tabcolsep}{-4pt}
\begin{sidewaystable}
\begin{tabular}{cc|rrrrrrrrrrrrrrrrrrrrrrrrrrrrrrrrrrrr}
$k$ & $j$: & \phantom{0}0 &  \phantom{0}1 &  \phantom{0}2 &  \phantom{0}3 &  \phantom{0}4 &  \phantom{0}5 &  \phantom{0}6 &  \phantom{0}7 &  \phantom{0}8 &  \phantom{0}9 & 10 & 11 & 12 & 13 & 14 & 15 & 16 & 17 & 18 & 19 & 20 & 21 & 22 & 23 & 24 & 25 & 26 & 27 & 28 & 39 & 30 & 31 & 32 & 33 & 34 & 35 \\ \hline
0 & & 1 & 1 & 1 & 1 & 1 & 1 & 1 & 1 & 1 & 1 & 1 & 1 & 1 & 1 & 1 & 1 & 1 & 1 & 1 & 1 & 1 & 1 & 1 & 1 & 1 & 1 & 1 & 1 & 1 & 1 & 1 & 1 & 1 & 1 & 1 & 1 \\
1 & & 1 & -1 & -1 & -1 & 1 & 1 & -1 & -1 & -1 & 1 & 1 & 1 & 1 & 1 & 1 & 1 & 1 & 1 & -1 & -1 & -1 & -1 & -1 & -1 & 1 & 1 & -1 & -1 & -1 & -1 & -1 & -1 & 1 & 1 & 1 & 1 \\
2 & & 1 & -1 & -1 & -1 & 1 & 1 & 1 & 1 & 1 & -1 & -1 & -1 & -1 & -1 & -1 & -1 & -1 & -1 & 1 & 1 & 1 & 1 & 1 & 1 & 1 & 1 & -1 & -1 & -1 & -1 & -1 & -1 & 1 & 1 & 1 & 1 \\
3 & & 1 & 1 & 1 & 1 & 1 & 1 & -1 & -1 & -1 & -1 & -1 & -1 & -1 & -1 & -1 & -1 & -1 & -1 & -1 & -1 & -1 & -1 & -1 & -1 & 1 & 1 & 1 & 1 & 1 & 1 & 1 & 1 & 1 & 1 & 1 & 1 \\
4 & & 4 & -4 & -4 & -4 & 4 & 4 & 0 & 0 & 0 & 0 & 0 & 0 & 0 & 0 & 0 & 0 & 0 & 0 & 0 & 0 & 0 & 0 & 0 & 0 & -2 & -2 & 2 & 2 & 2 & 2 & 2 & 2 & -2 & -2 & -2 & -2 \\
5 & & 4 & 4 & 4 & 4 & 4 & 4 & 0 & 0 & 0 & 0 & 0 & 0 & 0 & 0 & 0 & 0 & 0 & 0 & 0 & 0 & 0 & 0 & 0 & 0 & -2 & -2 & -2 & -2 & -2 & -2 & -2 & -2 & -2 & -2 & -2 & -2 \\
6 & & 4 & 0 & 0 & 0 & -2 & -2 & -4 & -4 & -4 & 0 & 0 & 0 & 0 & 0 & 0 & 0 & 0 & 0 & 2 & 2 & 2 & 2 & 2 & 2 & 4 & 4 & 0 & 0 & 0 & 0 & 0 & 0 & -2 & -2 & -2 & -2 \\
7 & & 4 & 0 & 0 & 0 & -2 & -2 & 4 & 4 & 4 & 0 & 0 & 0 & 0 & 0 & 0 & 0 & 0 & 0 & -2 & -2 & -2 & -2 & -2 & -2 & 4 & 4 & 0 & 0 & 0 & 0 & 0 & 0 & -2 & -2 & -2 & -2 \\
8 & & 4 & 0 & 0 & 0 & -2 & -2 & 0 & 0 & 0 & 4 & -2 & -2 & -2 & -2 & 4 & -2 & -2 & 4 & 0 & 0 & 0 & 0 & 0 & 0 & -2 & -2 & 0 & 0 & 0 & 0 & 0 & 0 & 4 & -2 & -2 & 4 \\
9 & & 4 & 0 & 0 & 0 & -2 & -2 & 0 & 0 & 0 & -4 & 2 & 2 & 2 & 2 & -4 & 2 & 2 & -4 & 0 & 0 & 0 & 0 & 0 & 0 & -2 & -2 & 0 & 0 & 0 & 0 & 0 & 0 & 4 & -2 & -2 & 4 \\
10 & & 8 & 0 & 0 & 0 & -4 & -4 & 0 & 0 & 0 & 0 & 0 & 0 & 0 & 0 & 0 & 0 & 0 & 0 & 0 & 0 & 0 & 0 & 0 & 0 & -4 & -4 & 0 & 0 & 0 & 0 & 0 & 0 & -4 & 8 & 8 & -4 \\
\end{tabular}
\vskip 1ex
\caption{The characters $\bar{\eta}_k(\sigma_j)$ appearing in the inversion formula~\eqref{schemeinv}.
\label{deltamat}}
\end{sidewaystable}
\addtolength{\tabcolsep}{4pt}

\begin{table}[H]
\renewcommand*{\arraystretch}{1.2}
\begin{center}
\begin{tabular}{|c|c|c|l|}
\hline
$Q_{0}$ & $S$ & $+$ & $S_0$\\ \hline
$Q_{1}$ & $S$ & $+$ & $S_1+S_2+S_3$\\ \cline{2-3}
$Q_{2}$ & \multirow{2}{*}{$E$} & $-$ & $S_2-S_3$\\
$Q_{3}$ & & $+$ & $2S_1-S_2-S_3$\\ \hline
$Q_{4}$ & $S$ & $+$ & $S_4+S_5+S_6+S_8+S_9+S_{10}$\\ \cline{2-3}
$Q_{5}$ & $A$ & $-$ & $S_4-S_5-S_6+S_8+S_9-S_{10}$\\ \cline{2-3}
$Q_{6}$ & \multirow{2}{*}{$E$} & $-$ & $-S_4+S_5+S_6+2S_8-S_9-2S_{10}$\\
$Q_{7}$ & & $+$ & $3S_4-3S_6+3S_8-3S_{10}$\\ \cline{2-3}
$Q_{8}$ & \multirow{2}{*}{$E'$} & $-$ & $S_4+S_5-S_6-S_9$\\
$Q_{9}$ & & $+$ & $S_4+S_5+S_6-2S_8+S_9-2S_{10}$\\ \cline{2-3}
$Q_{10}$ & $S'$ & $+$ & $S_7$\\ \hline
$Q_{11}$ & $S$ & $+$ & $S_{11}+S_{13}+S_{16}$\\ \cline{2-3}
$Q_{12}$ & \multirow{2}{*}{$E$} & $-$ & $S_{13}-S_{16}$\\
$Q_{13}$ & & $+$ & $2S_{11}-S_{13}-S_{16}$\\ \cline{2-3}
$Q_{14}$ & $S'$ & $+$ & $S_{12}+S_{14}+S_{15}$\\ \cline{2-3}
$Q_{15}$ & \multirow{2}{*}{$E'$} & $-$ & $S_{12}-S_{14}$\\
$Q_{16}$ & & $+$ & $S_{12}+S_{14}-2S_{15}$\\ \hline
$Q_{17}$ & $A$ & $-$ & $(S_{17}-S_{20}+S_{22})/3$\\ \cline{2-3}
$Q_{18}$ & \multirow{2}{*}{$\bar{E}$} & $-$ & $(-S_{17}+S_{20}+2S_{22})/3$\\
$Q_{19}$ & & $+$ & $(S_{17}+S_{20})/3$\\ \cline{2-3}
$Q_{20}$ & $S$ & $+$ & $S_{18}+S_{19}+S_{21}$\\ \cline{2-3}
$Q_{21}$ & \multirow{2}{*}{$E$} & $-$ & $S_{19}-S_{21}$\\
$Q_{22}$ & & $+$ & $2S_{18}-S_{19}-S_{21}$\\ \hline
$Q_{23}$ & $S$ & $+$ & $(S_{24}+S_{25}+S_{27}+S_{29}+S_{30}+S_{31})/3$ \\ \cline{2-3}
$Q_{24}$ & $A$ & $-$ & $(S_{24}-S_{25}-S_{27}+S_{29}+S_{30}-S_{31})/3$ \\ \cline{2-3}
$Q_{25}$ & \multirow{2}{*}{$E$} & $-$ & $(-S_{24}-2S_{25}+S_{27}+S_{30})/3$ \\
$Q_{26}$ & & $+$ & $(-S_{24}-S_{25}-S_{27}+2S_{29}-S_{30}+2S_{31})/3$ \\ \cline{2-3}
$Q_{27}$ & \multirow{2}{*}{$E'$} & $-$ & $(S_{24}-S_{25}-S_{27}-2S_{29}+S_{30}+2S_{31})/3$ \\
$Q_{28}$ & & $+$ & $(-S_{24}+S_{25}-S_{27}+S_{30})/3$ \\ \cline{2-3}
$Q_{29}$ & $S'$ & $+$ & $S_{28}$ \\ \cline{2-3}
$Q_{30}$ & $S''$ & $+$ & $(S_{23}+S_{26}+S_{32})/6$ \\ \cline{2-3}
$Q_{31}$ & \multirow{2}{*}{$E''$} & $-$ & $(-S_{26}+S_{32})/6$ \\
$Q_{32}$ & & $+$ & $(-2S_{23}+S_{26}+S_{32})/6$ \\ \hline
$Q_{33}$ & $A$ & $-$ & $(S_{33}-S_{34}+S_{35})/3$ \\ \cline{2-3} \cline{2-3}
$Q_{34}$ & \multirow{2}{*}{$\bar{E}$} & $-$ & $(-S_{33}+S_{34}+2S_{35})/3$ \\
$Q_{35}$ & & $+$ & $(S_{33}+S_{34})/3$ \\ \hline
\end{tabular}
\end{center}
\caption{Linear combinations of shapes classified by the representations of the group $\mathcal{S}_3^{xyz}$ ($Q$-basis in Appendix~\ref{altclass}). The second column gives the representation. Distinct copies of the same representation within blocks of the same total degree are denoted by primes. The third column gives the parity under the transposition $x\leftrightarrow y$. The $Q_i$ are normalized as polynomials over $\mathbb{Z}$ in the harmonic polynomials~\eqref{cvf}, with all common divisors of the coefficients divided out. The change of basis is orthogonal by rows.
\label{evalQ}}
\end{table}

\section{Matrix elements}\label{matrixel}

Here we provide explicit formulas for the matrices $M_i$, $i=4,\ldots, 10$, in the main text. The matrix elements of $M_4$ are
\begin{equation}
\begin{split}
M_4(1,1) &= x_{121}y_{121} + x_{211}y_{121} + x_{211}y_{211} , \\
M_4(1,2) &= -x_{121}y_{121} + x_{211}y_{211} , \\
M_4(1,3) &= x_{211}y_{121} + x_{121}y_{211} + x_{211}y_{211}, \\
M_4(1,4) &= x_{211}y_{121} - x_{121}y_{211}, \\
M_4(2,1) &= -x_{221}y_{121} - x_{212}y_{211} - x_{221}y_{211}, \\
M_4(2,2) &= -x_{212}y_{121} - x_{212}y_{211} - x_{221}y_{211}, \\
M_4(2,3) &= -x_{212}y_{121} - x_{221}y_{121} - x_{221}y_{211}, \\
M_4(2,4) &= -x_{212}y_{121} - x_{221}y_{121} - x_{212}y_{211}, \\
M_4(3,1) &= x_{121}y_{212} - x_{211}y_{221}, \\
M_4(3,2) &= -x_{121}y_{212} - x_{211}y_{212} - x_{211}y_{221}, \\
M_4(3,3) &= -x_{121}y_{212} - x_{121}y_{221} - x_{211}y_{221}, \\
M_4(3,4) &= x_{121}y_{212} + x_{211}y_{212} + x_{121}y_{221}, \\
M_4(4,1) &= x_{212}y_{212} + x_{212}y_{221} + x_{221}y_{221}, \\
M_4(4,2) &= x_{221}y_{212} + x_{212}y_{221} + x_{221}y_{221}, \\
M_4(4,3) &= -x_{212}y_{212} + x_{221}y_{221}, \\
M_4(4,4) &= -x_{221}y_{212} + x_{212}y_{221}.
\end{split}
\end{equation}

The matrix elements of $M_5$ are
\begin{equation}
\begin{split}
M_5(1,1) &= -x_{121}y_{121} + x_{211}y_{121} - 2x_{121}y_{211} - x_{211}y_{211}, \\ 
M_5(1,2) &= -x_{121}y_{121} - 2x_{211}y_{121} - 2x_{121}y_{211} - x_{211}y_{211}, \\ 
M_5(1,3) &= -2x_{121}y_{121} - x_{211}y_{121} - x_{121}y_{211} + x_{211}y_{211}, \\ 
M_5(1,4) &= -2x_{121}y_{121} - x_{211}y_{121} - x_{121}y_{211} - 2x_{211}y_{211}, \\ 
M_5(2,1) &= -2x_{212}y_{121} - x_{221}y_{121} - x_{212}y_{211} + x_{221}y_{211}, \\ 
M_5(2,2) &= x_{212}y_{121} + 2x_{221}y_{121} - x_{212}y_{211} + x_{221}y_{211}, \\ 
M_5(2,3) &= -x_{212}y_{121} + x_{221}y_{121} - 2x_{212}y_{211} - x_{221}y_{211}, \\ 
M_5(2,4) &= -x_{212}y_{121} + x_{221}y_{121} + x_{212}y_{211} + 2x_{221}y_{211}, \\ 
M_5(3,1) &= -x_{121}y_{212} - 2x_{211}y_{212} - 2x_{121}y_{221} - x_{211}y_{221}, \\ 
M_5(3,2) &= -x_{121}y_{212} + x_{211}y_{212} - 2x_{121}y_{221} - x_{211}y_{221}, \\ 
M_5(3,3) &= x_{121}y_{212} + 2x_{211}y_{212} - x_{121}y_{221} + x_{211}y_{221}, \\ 
M_5(3,4) &= x_{121}y_{212} - x_{211}y_{212} - x_{121}y_{221} - 2x_{211}y_{221}, \\ 
M_5(4,1) &= -x_{212}y_{212} - 2x_{221}y_{212} + x_{212}y_{221} - x_{221}y_{221}, \\ 
M_5(4,2) &= 2x_{212}y_{212} + x_{221}y_{212} + x_{212}y_{221} - x_{221}y_{221}, \\ 
M_5(4,3) &= x_{212}y_{212} + 2x_{221}y_{212} + 2x_{212}y_{221} + x_{221}y_{221}, \\ 
M_5(4,4) &= -2x_{212}y_{212} - x_{221}y_{212} - x_{212}y_{221} - 2x_{221}y_{221}.
\end{split}
\end{equation}

The matrix elements of $M_6$ are
\begin{equation}
\begin{split}
M_6(1,1) &= x_{121}z_{121} + x_{211}z_{121} + x_{211}z_{211},\\ 
M_6(1,2) &= -x_{121}z_{121} + x_{211}z_{211},\\ 
M_6(1,3) &= x_{211}z_{121} + x_{121}z_{211} + x_{211}z_{211},\\ 
M_6(1,4) &= x_{211}z_{121} - x_{121}z_{211},\\ 
M_6(2,1) &= -x_{221}z_{121} - x_{212}z_{211} - x_{221}z_{211},\\ 
M_6(2,2) &= -x_{212}z_{121} - x_{212}z_{211} - x_{221}z_{211},\\ 
M_6(2,3) &= -x_{212}z_{121} - x_{221}z_{121} - x_{221}z_{211},\\ 
M_6(2,4) &= -x_{212}z_{121} - x_{221}z_{121} - x_{212}z_{211},\\ 
M_6(3,1) &= x_{121}z_{212} - x_{211}z_{221},\\ 
M_6(3,2) &= -x_{121}z_{212} - x_{211}z_{212} - x_{211}z_{221},\\ 
M_6(3,3) &= -x_{121}z_{212} - x_{121}z_{221} - x_{211}z_{221},\\ 
M_6(3,4) &= x_{121}z_{212} + x_{211}z_{212} + x_{121}z_{221},\\
M_6(4,1) &= x_{212}z_{212} + x_{212}z_{221} + x_{221}z_{221},\\ 
M_6(4,2) &= x_{221}z_{212} + x_{212}z_{221} + x_{221}z_{221},\\ 
M_6(4,3) &= -x_{212}z_{212} + x_{221}z_{221},\\ 
M_6(4,4) &= -x_{221}z_{212} + x_{212}z_{221}.
\end{split}
\end{equation}

The matrix elements of $M_7$ are
\begin{equation}
\begin{split}
M_7(1,1) &= -x_{121}z_{121} + x_{211}z_{121} - 2x_{121}z_{211} - x_{211}z_{211},\\ 
M_7(1,2) &= -x_{121}z_{121} - 2x_{211}z_{121} - 2x_{121}z_{211} - x_{211}z_{211},\\ 
M_7(1,3) &= -2x_{121}z_{121} - x_{211}z_{121} - x_{121}z_{211} + x_{211}z_{211},\\ 
M_7(1,4) &= -2x_{121}z_{121} - x_{211}z_{121} - x_{121}z_{211} - 2x_{211}z_{211},\\ 
M_7(2,1) &= -2x_{212}z_{121} - x_{221}z_{121} - x_{212}z_{211} + x_{221}z_{211},\\ 
M_7(2,2) &= x_{212}z_{121} + 2x_{221}z_{121} - x_{212}z_{211} + x_{221}z_{211},\\ 
M_7(2,3) &= -x_{212}z_{121} + x_{221}z_{121} - 2x_{212}z_{211} - x_{221}z_{211},\\ 
M_7(2,4) &= -x_{212}z_{121} + x_{221}z_{121} + x_{212}z_{211} + 2x_{221}z_{211},\\
M_7(3,1) &= -x_{121}z_{212} - 2x_{211}z_{212} - 2x_{121}z_{221} - x_{211}z_{221},\\ 
M_7(3,2) &= -x_{121}z_{212} + x_{211}z_{212} - 2x_{121}z_{221} - x_{211}z_{221},\\ 
M_7(3,3) &= x_{121}z_{212} + 2x_{211}z_{212} - x_{121}z_{221} + x_{211}z_{221},\\ 
M_7(3,4) &= x_{121}z_{212} - x_{211}z_{212} - x_{121}z_{221} - 2x_{211}z_{221},\\
M_7(4,1) &= -x_{212}z_{212} - 2x_{221}z_{212} + x_{212}z_{221} - x_{221}z_{221},\\ 
M_7(4,2) &= 2x_{212}z_{212} + x_{221}z_{212} + x_{212}z_{221} - x_{221}z_{221},\\ 
M_7(4,3) &= x_{212}z_{212} + 2x_{221}z_{212} + 2x_{212}z_{221} + x_{221}z_{221},\\ 
M_7(4,4) &= -2x_{212}z_{212} - x_{221}z_{212} - x_{212}z_{221} - 2x_{221}z_{221}.
\end{split}
\end{equation}

The matrix elements of $M_8$ are
\begin{equation}
\begin{split}
M_8(1,1) & = y_{121}z_{121} + 2y_{211}z_{121} - y_{121}z_{211} + y_{211}z_{211}, \\
M_8(1,2) & = 2y_{121}z_{121} + y_{211}z_{121} + y_{121}z_{211} + 2y_{211}z_{211}, \\
M_8(1,3) & = -y_{121}z_{121} + y_{211}z_{121} + y_{121}z_{211} + 2y_{211}z_{211}, \\
M_8(1,4) & = 2y_{121}z_{121} + y_{211}z_{121} + y_{121}z_{211} - y_{211}z_{211}, \\
M_8(2,1) & = -y_{212}z_{121} - 2y_{221}z_{121} - 2y_{212}z_{211} - y_{221}z_{211}, \\
M_8(2,2) & = y_{212}z_{121} - y_{221}z_{121} - y_{212}z_{211} - 2y_{221}z_{211}, \\
M_8(2,3) & = -2y_{212}z_{121} - y_{221}z_{121} - y_{212}z_{211} - 2y_{221}z_{211}, \\
M_8(2,4) & = y_{212}z_{121} - y_{221}z_{121} + 2y_{212}z_{211} + y_{221}z_{211}, \\
M_8(3,1) & = 2y_{121}z_{212} + y_{211}z_{212} + y_{121}z_{221} - y_{211}z_{221}, \\
M_8(3,2) & = y_{121}z_{212} - y_{211}z_{212} - y_{121}z_{221} - 2y_{211}z_{221}, \\
M_8(3,3) & = -2y_{121}z_{212} - y_{211}z_{212} - y_{121}z_{221} - 2y_{211}z_{221}, \\
M_8(3,4) & = y_{121}z_{212} + 2y_{211}z_{212} - y_{121}z_{221} + y_{211}z_{221}, \\
M_8(4,1) & = y_{212}z_{212} - y_{221}z_{212} + 2y_{212}z_{221} + y_{221}z_{221}, \\
M_8(4,2) & = 2y_{212}z_{212} + y_{221}z_{212} + y_{212}z_{221} + 2y_{221}z_{221}, \\
M_8(4,3) & = -y_{212}z_{212} + y_{221}z_{212} + y_{212}z_{221} + 2y_{221}z_{221}, \\
M_8(4,4) & = -y_{212}z_{212} - 2y_{221}z_{212} - 2y_{212}z_{221} - y_{221}z_{221}.
\end{split}
\end{equation}

The matrix elements of $M_9$ are
\begin{equation}
\begin{split}
M_9(1,1) & = y_{121}z_{121} + y_{211}z_{121} + y_{211}z_{211}, \\
M_9(1,2) & = -y_{121}z_{121} + y_{211}z_{211}, \\
M_9(1,3) & = y_{211}z_{121} + y_{121}z_{211} + y_{211}z_{211}, \\
M_9(1,4) & = y_{211}z_{121} - y_{121}z_{211}, \\
M_9(2,1) & = -y_{221}z_{121} - y_{212}z_{211} - y_{221}z_{211}, \\
M_9(2,2) & = -y_{212}z_{121} - y_{212}z_{211} - y_{221}z_{211}, \\
M_9(2,3) & = -y_{212}z_{121} - y_{221}z_{121} - y_{221}z_{211}, \\
M_9(2,4) & = -y_{212}z_{121} - y_{221}z_{121} - y_{212}z_{211}, \\
M_9(3,1) & = -y_{121}z_{212} + y_{211}z_{221}, \\
M_9(3,2) & = y_{121}z_{212} + y_{211}z_{212} + y_{211}z_{221}, \\
M_9(3,3) & = y_{121}z_{212} + y_{121}z_{221} + y_{211}z_{221}, \\
M_9(3,4) & = -y_{121}z_{212} - y_{211}z_{212} - y_{121}z_{221},\\
M_9(4,1) & = y_{212}z_{212} + y_{212}z_{221} + y_{221}z_{221}, \\
M_9(4,2) & = y_{221}z_{212} + y_{212}z_{221} + y_{221}z_{221}, \\
M_9(4,3) & = -y_{212}z_{212} + y_{221}z_{221}, \\
M_9(4,4) & = -y_{221}z_{212} + y_{212}z_{221}.
\end{split}
\end{equation}

The matrix elements of $M_{10}$ are
\begin{equation}
\begin{split}
M_{10}(1,1) &= -2x_{121}y_{121}z_{121} - x_{121}y_{211}z_{121} - x_{121}y_{121}z_{211} + x_{121}y_{211}z_{211},\\ 
M_{10}(1,2) &= x_{121}y_{121}z_{121} + 2x_{121}y_{211}z_{121} - x_{121}y_{121}z_{211} + x_{121}y_{211}z_{211},\\ 
M_{10}(1,3) &= -x_{121}y_{121}z_{121} + x_{121}y_{211}z_{121} - 2x_{121}y_{121}z_{211} - x_{121}y_{211}z_{211},\\ 
M_{10}(1,4) &= -x_{121}y_{121}z_{121} + x_{121}y_{211}z_{121} + x_{121}y_{121}z_{211} + 2x_{121}y_{211}z_{211},\\ 
M_{10}(1,5) &= x_{121}y_{121}z_{121} + x_{211}y_{121}z_{121} + x_{121}y_{211}z_{121} \\
&\phantom{x_{121}y_{121}z_{121} + x_{211}} + x_{211}y_{211}z_{121} + x_{121}y_{211}z_{211} + x_{211}y_{211}z_{211},\\ 
M_{10}(1,6) &= x_{121}y_{121}z_{121} + x_{211}y_{121}z_{121} + x_{121}y_{121}z_{211} \\
&\phantom{x_{121}y_{121}z_{121} + x_{211}} + x_{211}y_{121}z_{211} + x_{121}y_{211}z_{211} + x_{211}y_{211}z_{211},\\ 
M_{10}(1,7) &= -x_{121}y_{121}z_{121} - x_{211}y_{121}z_{121} - x_{121}y_{211}z_{121} \\
&\phantom{x_{121}y_{121}z_{121} + x_{211}} - x_{211}y_{211}z_{121} - x_{121}y_{121}z_{211} - x_{211}y_{121}z_{211},\\ 
M_{10}(1,8) &= -x_{121}y_{121}z_{121} - x_{211}y_{121}z_{121} + x_{121}y_{211}z_{211} + x_{211}y_{211}z_{211},\\
M_{10}(2,1) &= -2x_{212}y_{121}z_{121} - x_{212}y_{211}z_{121} - x_{212}y_{121}z_{211} + x_{212}y_{211}z_{211},\\ 
M_{10}(2,2) &= x_{212}y_{121}z_{121} + 2x_{212}y_{211}z_{121} - x_{212}y_{121}z_{211} + x_{212}y_{211}z_{211},\\ 
M_{10}(2,3) &= -x_{212}y_{121}z_{121} + x_{212}y_{211}z_{121} - 2x_{212}y_{121}z_{211} - x_{212}y_{211}z_{211},\\ 
M_{10}(2,4) &= -x_{212}y_{121}z_{121} + x_{212}y_{211}z_{121} + x_{212}y_{121}z_{211} + 2x_{212}y_{211}z_{211},\\ 
M_{10}(2,5) &= -x_{221}y_{121}z_{121} - x_{221}y_{211}z_{121} - x_{221}y_{211}z_{211},\\ 
M_{10}(2,6) &= -x_{221}y_{121}z_{121} - x_{221}y_{121}z_{211} - x_{221}y_{211}z_{211},\\ 
M_{10}(2,7) &= x_{221}y_{121}z_{121} + x_{221}y_{211}z_{121} + x_{221}y_{121}z_{211},\\ 
M_{10}(2,8) &= x_{221}y_{121}z_{121} - x_{221}y_{211}z_{211},\\ 
M_{10}(3,1) &= -x_{121}y_{212}z_{121} + x_{121}y_{221}z_{121} - 2x_{121}y_{212}z_{211} - x_{121}y_{221}z_{211},\\ 
M_{10}(3,2) &= -x_{121}y_{212}z_{121} - 2x_{121}y_{221}z_{121} - 2x_{121}y_{212}z_{211} - x_{121}y_{221}z_{211},\\ 
M_{10}(3,3) &= -2x_{121}y_{212}z_{121} - x_{121}y_{221}z_{121} - x_{121}y_{212}z_{211} + x_{121}y_{221}z_{211},\\ 
M_{10}(3,4) &= -2x_{121}y_{212}z_{121} - x_{121}y_{221}z_{121} - x_{121}y_{212}z_{211} - 2x_{121}y_{221}z_{211},\\ 
M_{10}(3,5) &= -x_{121}y_{221}z_{121} - x_{211}y_{221}z_{121} - x_{121}y_{212}z_{211} \\
&\phantom{x_{121}y_{121}z_{121} + x_{211}} - x_{211}y_{212}z_{211} - x_{121}y_{221}z_{211} - x_{211}y_{221}z_{211},\\ 
M_{10}(3,6) &= x_{121}y_{212}z_{121} + x_{211}y_{212}z_{121} - x_{121}y_{221}z_{211} - x_{211}y_{221}z_{211},\\ 
M_{10}(3,7) &= x_{121}y_{221}z_{121} + x_{211}y_{221}z_{121} - x_{121}y_{212}z_{211} - x_{211}y_{212}z_{211},\\ 
M_{10}(3,8) &= -x_{121}y_{212}z_{121} - x_{211}y_{212}z_{121} - x_{121}y_{212}z_{211} \\
&\phantom{x_{121}y_{121}z_{121} + x_{211}} - x_{211}y_{212}z_{211} - x_{121}y_{221}z_{211} - x_{211}y_{221}z_{211},\\ 
M_{10}(4,1) &= -x_{121}y_{121}z_{212} - 2x_{121}y_{211}z_{212} + x_{121}y_{121}z_{221} - x_{121}y_{211}z_{221},\\ 
M_{10}(4,2) &= 2x_{121}y_{121}z_{212} + x_{121}y_{211}z_{212} + x_{121}y_{121}z_{221} - x_{121}y_{211}z_{221},\\ 
M_{10}(4,3) &= x_{121}y_{121}z_{212} + 2x_{121}y_{211}z_{212} + 2x_{121}y_{121}z_{221} + x_{121}y_{211}z_{221},\\ 
M_{10}(4,4) &= -2x_{121}y_{121}z_{212} - x_{121}y_{211}z_{212} - x_{121}y_{121}z_{221} - 2x_{121}y_{211}z_{221},\\ 
M_{10}(4,5) &= x_{121}y_{121}z_{212} + x_{211}y_{121}z_{212} - x_{121}y_{211}z_{221} - x_{211}y_{211}z_{221},\\ 
M_{10}(4,6) &= -x_{121}y_{211}z_{212} - x_{211}y_{211}z_{212} - x_{121}y_{121}z_{221} \\
&\phantom{x_{121}y_{121}z_{121} + x_{211}} - x_{211}y_{121}z_{221} - x_{121}y_{211}z_{221} - x_{211}y_{211}z_{221},\\ 
M_{10}(4,7) &= -x_{121}y_{211}z_{212} - x_{211}y_{211}z_{212} + x_{121}y_{121}z_{221} + x_{211}y_{121}z_{221},\\ 
M_{10}(4,8) &= -x_{121}y_{121}z_{212} - x_{211}y_{121}z_{212} - x_{121}y_{211}z_{212} \\
&\phantom{x_{121}y_{121}z_{121} + x_{211}} - x_{211}y_{211}z_{212} - x_{121}y_{211}z_{221} - x_{211}y_{211}z_{221},\\ 
\end{split}
\label{first}
\end{equation}

\begin{equation*}\tag{\ref{first}}
\begin{split}
M_{10}(5,1) &= x_{212}y_{212}z_{121} - x_{212}y_{221}z_{121} + 2x_{212}y_{212}z_{211} + x_{212}y_{221}z_{211},\\ 
M_{10}(5,2) &= x_{212}y_{212}z_{121} + 2x_{212}y_{221}z_{121} + 2x_{212}y_{212}z_{211} + x_{212}y_{221}z_{211},\\ 
M_{10}(5,3) &= 2x_{212}y_{212}z_{121} + x_{212}y_{221}z_{121} + x_{212}y_{212}z_{211} - x_{212}y_{221}z_{211},\\ 
M_{10}(5,4) &= 2x_{212}y_{212}z_{121} + x_{212}y_{221}z_{121} + x_{212}y_{212}z_{211} + 2x_{212}y_{221}z_{211},\\ 
M_{10}(5,5) &= -x_{221}y_{221}z_{121} - x_{221}y_{212}z_{211} - x_{221}y_{221}z_{211},\\ 
M_{10}(5,6) &= x_{221}y_{212}z_{121} - x_{221}y_{221}z_{211},\\ 
M_{10}(5,7) &= x_{221}y_{221}z_{121} - x_{221}y_{212}z_{211},\\ 
M_{10}(5,8) &= -x_{221}y_{212}z_{121} - x_{221}y_{212}z_{211} - x_{221}y_{221}z_{211},\\ 
M_{10}(6,1) &= x_{212}y_{121}z_{212} + 2x_{212}y_{211}z_{212} - x_{212}y_{121}z_{221} + x_{212}y_{211}z_{221},\\ 
M_{10}(6,2) &= -2x_{212}y_{121}z_{212} - x_{212}y_{211}z_{212} - x_{212}y_{121}z_{221} + x_{212}y_{211}z_{221},\\ 
M_{10}(6,3) &= -x_{212}y_{121}z_{212} - 2x_{212}y_{211}z_{212} - 2x_{212}y_{121}z_{221} - x_{212}y_{211}z_{221},\\ 
M_{10}(6,4) &= 2x_{212}y_{121}z_{212} + x_{212}y_{211}z_{212} + x_{212}y_{121}z_{221} + 2x_{212}y_{211}z_{221},\\ 
M_{10}(6,5) &= x_{221}y_{121}z_{212} - x_{221}y_{211}z_{221},\\ 
M_{10}(6,6) &= -x_{221}y_{211}z_{212} - x_{221}y_{121}z_{221} - x_{221}y_{211}z_{221},\\ 
M_{10}(6,7) &= -x_{221}y_{211}z_{212} + x_{221}y_{121}z_{221},\\ 
M_{10}(6,8) &= -x_{221}y_{121}z_{212} - x_{221}y_{211}z_{212} - x_{221}y_{211}z_{221},\\ 
M_{10}(7,1) &= -x_{121}y_{212}z_{212} - 2x_{121}y_{221}z_{212} - 2x_{121}y_{212}z_{221} - x_{121}y_{221}z_{221},\\ 
M_{10}(7,2) &= -x_{121}y_{212}z_{212} + x_{121}y_{221}z_{212} - 2x_{121}y_{212}z_{221} - x_{121}y_{221}z_{221},\\ 
M_{10}(7,3) &= x_{121}y_{212}z_{212} + 2x_{121}y_{221}z_{212} - x_{121}y_{212}z_{221} + x_{121}y_{221}z_{221},\\ 
M_{10}(7,4) &= x_{121}y_{212}z_{212} - x_{121}y_{221}z_{212} - x_{121}y_{212}z_{221} - 2x_{121}y_{221}z_{221},\\ 
M_{10}(7,5) &= -x_{121}y_{212}z_{212} - x_{211}y_{212}z_{212} - x_{121}y_{212}z_{221} \\
&\phantom{x_{121}y_{121}z_{121} + x_{211}} - x_{211}y_{212}z_{221} - x_{121}y_{221}z_{221} - x_{211}y_{221}z_{221},\\ 
M_{10}(7,6) &= -x_{121}y_{212}z_{212} - x_{211}y_{212}z_{212} - x_{121}y_{221}z_{212} \\
&\phantom{x_{121}y_{121}z_{121} + x_{211}} - x_{211}y_{221}z_{212} - x_{121}y_{221}z_{221} - x_{211}y_{221}z_{221},\\ 
M_{10}(7,7) &= -x_{121}y_{212}z_{212} - x_{211}y_{212}z_{212} - x_{121}y_{221}z_{212} \\
&\phantom{x_{121}y_{121}z_{121} + x_{211}} - x_{211}y_{221}z_{212} - x_{121}y_{212}z_{221} - x_{211}y_{212}z_{221},\\ 
M_{10}(7,8) &= -x_{121}y_{221}z_{212} - x_{211}y_{221}z_{212} - x_{121}y_{212}z_{221} \\
&\phantom{x_{121}y_{121}z_{121} + x_{211}} - x_{211}y_{212}z_{221} - x_{121}y_{221}z_{221} - x_{211}y_{221}z_{221},\\ 
M_{10}(8,1) &= -x_{212}y_{212}z_{212} - 2x_{212}y_{221}z_{212} - 2x_{212}y_{212}z_{221} - x_{212}y_{221}z_{221},\\ 
M_{10}(8,2) &= -x_{212}y_{212}z_{212} + x_{212}y_{221}z_{212} - 2x_{212}y_{212}z_{221} - x_{212}y_{221}z_{221},\\ 
M_{10}(8,3) &= x_{212}y_{212}z_{212} + 2x_{212}y_{221}z_{212} - x_{212}y_{212}z_{221} + x_{212}y_{221}z_{221},\\ 
M_{10}(8,4) &= x_{212}y_{212}z_{212} - x_{212}y_{221}z_{212} - x_{212}y_{212}z_{221} - 2x_{212}y_{221}z_{221},\\ 
M_{10}(8,5) &= x_{221}y_{212}z_{212} + x_{221}y_{212}z_{221} + x_{221}y_{221}z_{221},\\ 
M_{10}(8,6) &= x_{221}y_{212}z_{212} + x_{221}y_{221}z_{212} + x_{221}y_{221}z_{221},\\ 
M_{10}(8,7) &= x_{221}y_{212}z_{212} + x_{221}y_{221}z_{212} + x_{221}y_{212}z_{221},\\ 
M_{10}(8,8) &= x_{221}y_{221}z_{212} + x_{221}y_{212}z_{221} + x_{221}y_{221}z_{221}.
\end{split}
\end{equation*}

\section{Alternative classifications of shapes}\label{altclass}

In the main text, we have applied the theory of shapes in their original presentation as iterated derivatives, characterized by the space directions $x$, $y$, and $z$, as listed in Table~\ref{evalS} (the $S$-basis). Because the three directions are equivalent, it is a natural alternative to present the shapes as spanning representations of the permutation group of the three coordinate axes, denoted $\mathcal{S}_3^{xyz}$. This program is easy to carry out, because $\mathcal{S}_3^{xyz}$ cannot change the degree of an expression, so one is limited to the small subspaces of shapes of the same degree, further reduced by grouping wave functions belonging to the blocks $I_k$, according to the analysis above.

For example, $\mathcal{S}_3^{xyz}$ connects shapes of the same degree belonging to blocks $I_4$, $I_6$, and $I_8$. Referring to Table~\ref{evalS}, one is led to consider separately the triplet $(S_1, S_2, S_3)$, the triplet $(S_{11}, S_{13}, S_{16})$, and the sextuplet $(S_4, S_5, S_6, S_8, S_9, S_{10})$. The former two contain the symmetric ($S$) and two-dimensional ($E$) representations of $\mathcal{S}_3^{xyz}$, while the sextuplet splits into one $S$, one antisymmetric ($A$), and two $E$'s. In this way, one finds a linear transformation from the $S$-basis to the basis classified by representations of $\mathcal{S}_3^{xyz}$ (the $Q$-basis), listed in Table~\ref{evalQ}.

A remarkable feature of the $Q$-basis is that it distinguishes triplets containing the factors $x_{210}=y_{210}=z_{210}=-1$, which break down into an space-\emph{antisymmetric} planar singlet and a distinct (``left-handed'') copy of the representation $E$, denoted by $\bar{E}$ in Appendix~\ref{conventab}. These triplets are analogous to axial vectors in angular-momentum theory, which begs the question whether it is possible to present shapes as angular momentum multiplets as well.

The surprising answer is: Not in general. The situation can be understood already in the simplest case of two fermions~\cite{Rozman20}. Referring to the decomposition~\eqref{2part}, it is obvious that $(x,y,z)$ are three components of a polar vector, which comprises the lowest-degree triplet. But what is $xyz$? A perusal of tables of solid harmonics $\mathcal{Y}_{lm}$ identifies it as
\begin{equation}
xyz \propto\left(\mathcal{Y}_{32}-\mathcal{Y}_{3,-2}\right).
\end{equation} 
Thus, the shape $xyz$ needs to be complemented with bosonic excitations of the lowest-degree triplet $(x,y,z)$ in order to build a complete multiplet of good angular momentum $L=3$.

The same situation is observed with three fermions. There are only three pure-shape angular momentum multiplets, the ground-state triplet $(S_{33}, S_{34}, S_{35})$, and another triplet ($L=1$) and septiplet ($L=3$) spanned by the ten shapes $S_{23},\ldots,S_{32}$. The highest-projection state $L=M=3$ of the septiplet reads, up to normalization,
\begin{equation}
\begin{split}
[&2S_{29}+S_{32}-i(2S_{31}+S_{26})]/3
=\\
&=
[x_1 - x_2 + i(y_1 - y_2)][x_2 - x_3 + i(y_2 - y_3)]
[x_1 - x_3 + i(y_1 - y_3)],
\end{split}
\end{equation} 
from which all the other states within the same span can be easily reconstructed. The contrast between the partial incompatibility of shapes with the continuous group of rotations, and their perfect compatibility with the finite permutation group $\mathcal{S}_3^{xyz}$, has fundamental connotations that we elaborate in the Discussion.

We understand why the lowest-degree shapes and the shapes one degree higher span closed multiplets of the rotation group. Bosonic excitations of three particles are symmetric functions, built from the components (elementary symmetric functions~\cite{Stanley99}) $x_1+x_2+x_3$, $x_1x_2 + x_2x_3 + x_1x_3$, and $x_1x_2x_3$ (similarly for $y$ and $z$). Of these, the only linear term is the center-of-mass coordinate $x_1+x_2+x_3$, which cannot describe internal excitations. The lowest-degree symmetric functions describing relative motion are quadratic, e.g., a sum of terms $(x_i-x_j)^2$. Hence, one cannot dress a shape in bosonic excitations at a degree that is immediately above the lowest one. Because the complete Hilbert space is rotationally invariant, these shapes of low degree must realize that invariance without the help of bosonic functions.

\end{document}